\documentclass[sigconf]{acmart}
\usepackage{flushend}
\usepackage{amsmath}
\usepackage{booktabs} 
\usepackage[ruled]{algorithm2e} 

\SetAlFnt{\small}
\SetAlCapFnt{\small}
\SetAlCapNameFnt{\small}
\SetAlCapHSkip{0pt}
\IncMargin{-\parindent}

\renewcommand\footnotetextcopyrightpermission[1]{} 
\acmConference[arXiv.org]{Feb}{2020}{}
\acmYear{2020}
\copyrightyear{2018}

\setcopyright{usgovmixed}

\acmDOI{0000001.0000001}

\begin{document}
\title{EyeTAP: A Novel Technique using Voice Inputs to Address the Midas Touch Problem for Gaze-based Interactions}

\author{Mohsen Parisay}
\affiliation{%
  \institution{Concordia University}
  \streetaddress{1515 Saint-Catherine West}
  \city{Montreal}
  \state{Quebec}
  \postcode{H3G 2W1}
  \country{Canada}}
\email{m\_parisa@encs.concordia.ca}

\author{Charalambos Poullis}
\affiliation{%
  \institution{Concordia University}
  \city{Montreal}
  \state{Quebec}
  \country{Canada}
}
\email{charalambos@poullis.org}

\author{Marta Kersten}
\affiliation{%
 \institution{Concordia University}
 \streetaddress{1515 Saint-Catherine West}
 \city{Montreal}
 \state{Quebec}
 \country{Canada}}
\email{marta.kersten@concordia.ca}

\begin{abstract}

One of the main challenges of gaze-based interactions is the ability to distinguish normal eye function from a deliberate interaction with the computer system, commonly referred to as 'Midas touch'. In this paper we propose, EyeTAP (Eye tracking point-and-select by Targeted Acoustic Pulse) a hands-free interaction method for point-and-select tasks. We evaluated the prototype in two separate user studies, each containing two experiments with 33 participants and found that EyeTAP is robust even in presence of ambient noise in the audio input signal with tolerance of up to 70 dB, results in a faster movement time, and faster task completion time, and has a lower cognitive workload than voice recognition. In addition, EyeTAP has a lower error rate than the dwell-time method in a ribbon-shaped experiment. These characteristics make it applicable for users for whom physical movements are restricted or not possible due to a disability. Furthermore, EyeTAP has no specific requirements in terms of user interface design and therefore it can be easily integrated into existing systems with minimal modifications. EyeTAP can be regarded as an acceptable alternative to address the Midas touch.

\end{abstract}

%
%


%
%



\maketitle


\section{Introduction}

Modern eye tracking sensors offer a suitable alternative to  conventional input devices (i.e. keyboard and mouse) for users for whom manual interaction might be difficult or impossible. However, gaze-based interaction has well-known challenges the most important of which are (1) \textit{Midas touch} where a system cannot distinguish the basic function of the eye (i.e. looking and perceiving) from deliberate interaction with the system, and (2) \textit{eye jitter} which is caused by small physiological eye movements occurring during a fixation to perceive a scene visually \cite{JitterDefinition}. In this paper, we propose EyeTAP (Eye tracking point-and-select by Targeted Acoustic Pulse), an effective multimodal solution to the Midas touch problem. Specifically, our method integrates the user's gaze to control the mouse with audio input captured using a microphone to trigger button-press events for real-time interaction.

The contributions of this paper are twofold. Firstly, we have designed and developed an effective, multimodal interaction technique EyeTAP. The proposed approach is low-cost and allows for a completely hands-free interaction solution between the user and the computer system using only an eye-tracker and an audio input device. Secondly, we present two independent user studies each with two experiments comparing EyeTAP with all other widely-used interaction techniques. The analysis of the results clearly shows that using EyeTAP has at least comparable performance with the mouse. Furthermore, EyeTAP reaches competitive performance with the remaining eye-based interaction methods in cases where users would have restricted physical movement, or where manual interaction with an input device is not possible, e.g. medical practitioner having both hands busy. 


\section{Related Work}

In eye-based interaction, the Midas touch problem occurs when a user accidentally activates a computer command by looking when the intention was simply to look around and perceive the scene. According to Jacob \cite{MidasTouchDefinition}, this problem occurs because eye movements are natural, e.g. the eyes are used to look around an object or to scan a scene, often without any intention to activate a command or function.  This phenomenon is one of the major challenges in eye interaction techniques and diverse methods have been proposed to address the Midas touch problem. The solutions can be categorized into four groups according to the interaction technique they employ: (a) dwell-time processing, (b) smooth pursuits, (c) gaze gestures, and (d) multimodal interaction. Below, we describe each of these solutions and provide example use-cases. 

\subsection{Dwell-time processing} 
Dwell-time is the amount of time that the eye gaze must remain on a specific target in order to trigger an event. Researchers have tried to detect specific thresholds to handle the Midas touch problem \cite{ProbabilisticDwellTime, FocalFixations}. For example, Pi \emph{et al.} proposed a probabilistic model for text entry using eye gaze \cite{ProbabilisticDwellTime}. They reduced the Midas touch problem by assigning each letter a probability value based on the previously chosen letter such that a letter with lower probability requires a longer activation time to be activated and vice-versa. Velichkovsky \emph{et al.} applied focal fixations to resolve the Midas touch problem by assigning the mean duration time (empirically set to 325 ms) of a visual search task to trigger a function \cite{FocalFixations}. 

Dwell time has been shown to be even faster than the mouse in certain tasks, e.g. selecting a letter given an auditory cue~\cite{SibertJacob2000}. However, with dwell time there is a trade-off between accuracy and speed \cite{10.1145/1028014.1028045, 10.1145/1452392.1452443, majaranta2006effects}. The method of applying focal fixations may be very subjective since searching time varies across users \cite{Bednarik_Gowases_Tukiainen_2009}. Moreover, increasing the threshold may increase the duration time of the entire interaction. Conversely, reducing the amount of dwell-time may lead to more errors for some users \cite{10.1145/1452392.1452443}.


\subsection{Smooth pursuits} 
Smooth pursuits are a form of eye movement that occurs when a moving stimulus (e.g. an object or animation) is followed with gaze \cite{Barnes2012}. The method is typically implemented by using two visual points on the interface that appear on top and below each target. Then to activate the target the user must fixate on one of these points. This technique has been used to select targets~\cite{Pursuits}, control home appliances~\cite{AmbiGaze}, to activate functions such as mouse clicks~\cite{GazeEverywhere} or to use the music player on a smartwatch~\cite{OrbitsSmartWatch}. Schenk \emph{et al.} proposed a framework (GazeEverywhere) which enables users to replace mouse inputs \cite{GazeEverywhere}. This solution includes a computer to process gaze interactions (gaze PC),  a computer to show the results (unmodified PC) which are connected via a micro-controller to trigger mouse click events, and a glass pane to project gaze targets on a second screen. 

Vidal \emph{et al.} introduced an interaction technique (Pursuits) for large screens using moving objects to be activated by eye gaze \cite{Pursuits}. They used a Tobii X300 eye tracker and a public display to select targets on the screen. Velloso \emph{et al.} presented a framework (AmbiGaze) to control ambient devices such as TVs and stereos (each assigned with an infrared (IR) beacon) with eye gaze using a head-mounted eye tracker \cite{AmbiGaze}. The system employs a server to process gaze inputs and control the devices. Esteves \emph{et al.} presented a framework for a multi-touch Android smartwatch (Callisto 300) to input commands using a head-mounted eye tracker (Pupil Pro) \cite{OrbitsSmartWatch}. They developed three use-cases: a music player, a notifications panel with six colored points on the smartwatch screen representing six applications (e.g. social media apps), and a missed call menu with four commands, call back, reply text, save number and clear the notification.

	
\subsection{Gaze gestures} 
Gaze gestures are sequences of eye movements that follow a predefined pattern in a specific order~\cite{GazeGesture2007}. Researchers have proposed techniques which can be applied to analyze eye movements to detect unique gestures (e.g. ~\cite{GazeGesture2016, GazeGesture2007, GazeGesture2012, GazeGesture2010}). Drewes \emph{et al.} assigned up, down, left, right and diagonal directions to different characters on the keyboard thereby allowing a user to select a letter by moving the eye gaze in any direction \cite{GazeGesture2007}. In addition, they tried to distinguish between natural and intentional eye movements by using short fixation times during gesture detection and long fixation times to reset the gesture recognition. Istance \emph{et al.} developed two-legged and three-legged gaze gestures (up, down and diagonal patterns) for command selection to play World of Warcraft for users with motor impairment disabilities \cite{GazeGesture2010}. 

In a similar work, Hyrskykari \emph{et al.} studied both dwell-time and gaze gesture interactions in the context of video games and found that gaze gestures had better performance for command activation \cite{GazeGesture2012}. Moreover, gaze gestures produced fewer errors than the dwell-time and led to less visual distractions. B{\^a}ce \emph{et al.} proposed an AR prototype, containing a head-mounted eye tracker and a smartwatch, to embed virtual messages to real-world objects to be shared with peer users~\cite{GazeGesture2016}. The authors integrated eye gaze gestures as a pattern to encode and decode messages attached to a specific object previously tagged by another peer user, thus using gaze gestures as an authentication mechanism for secure communication.

	
\subsection{Multimodal Interaction} 
Multimodal techniques apply extra inputs from another modality (e.g. touch, audio, etc.) as the trigger of a function in addition to eye tracking. They can be divided into the following sub-categories: using mechanical switches, touch interaction, or facial gestures. 

\subsubsection{Applying a specific (mechanical) switch}	
For some specific domains, such as rehabilitation, and user groups (i.e. users with motor impairments or severe disabilities), researchers have applied specific switches to activate an event or function. For instance, Rajanna \emph{et al.} proposed a combined framework for users with disabilities which applies a foot pedal device to click on objects and to enter text~\cite{FootSwitch}. Meena \emph{et al.} applied a soft button on a wheelchair to control the movements of the wheelchair in different directions (horizontal, vertical and diagonal)~\cite{WheelchairSwitch}. Sidorakis \emph{et al.} applied a switch for a gazed-controlled multimedia framework on virtual reality head-mounted displays (Oculus Rift) to resolve the Midas touch problem~\cite{BinocularEyeTracking}. Biswas \emph{et al.} proposed a joystick to control point-and-select tasks for combat aviation platforms to address the Midas touch problem~\cite{JoystickSwitch}.

				
\subsubsection{Touch interaction} 
Some researchers have proposed the integration of using touch interaction, for a limited number of functions, to increase the accuracy of target selection. Pfeuffer \emph{et al.} applied a cursor at the gaze point to be controlled by a finger holding a tablet where a finger tap on the screen leads to a click on the current location of the pointer (CursorShift method)~\cite{GazeAndTouch01}. In a similar study by Pfeuffer \emph{et al.}, the authors investigated the integration of finger touch and pen inputs on a tablet for zooming or annotating tasks on images \cite{GazeAndTouch02}. Although this technique was not introduced as a solution to the Midas touch problem, it can increase the accuracy of selection which leads to reducing Midas touch. 


\subsubsection{Facial gestures recognition} 
In \cite{FaceGestures}, Rozado \emph{et al.} studied the potential of using live video monitoring to detect facial gestures to enhance eye tracking interaction. In their work (FaceSwitch), they associated facial gestures (opening mouth, raising eyebrows, smiling and twitching the nose up and down) to simulate left and right mouse clicks and customized some keyboard functions such as page down key press. 




Using a multimodal solution that combines eye-gaze with acoustic inputs (audio or speech detection) can be regarded as an alternative to the reviewed solutions and has the advantage of not requiring either extra hardware or a specialized user interface design. For this reason, we designed EyeTAP to use audio processing for selection. Our solution: (1) provides a hands-free interaction technique for users with special needs, and (2) addresses the Midas touch problem. Although there has been some work done on audio detection to simulate system events for computer interactions (e.g.~\cite{Blui, Cappella,DirectManipulationPatternRecognition}) the focus has been on signal processing for complex interactions. Conversely, in our work we applied acoustic inputs only as a way of sending commands. 


\section{EyeTAP Prototype}

\begin{figure}[htbp]
	\centering
		\includegraphics[width=0.30 \textwidth]{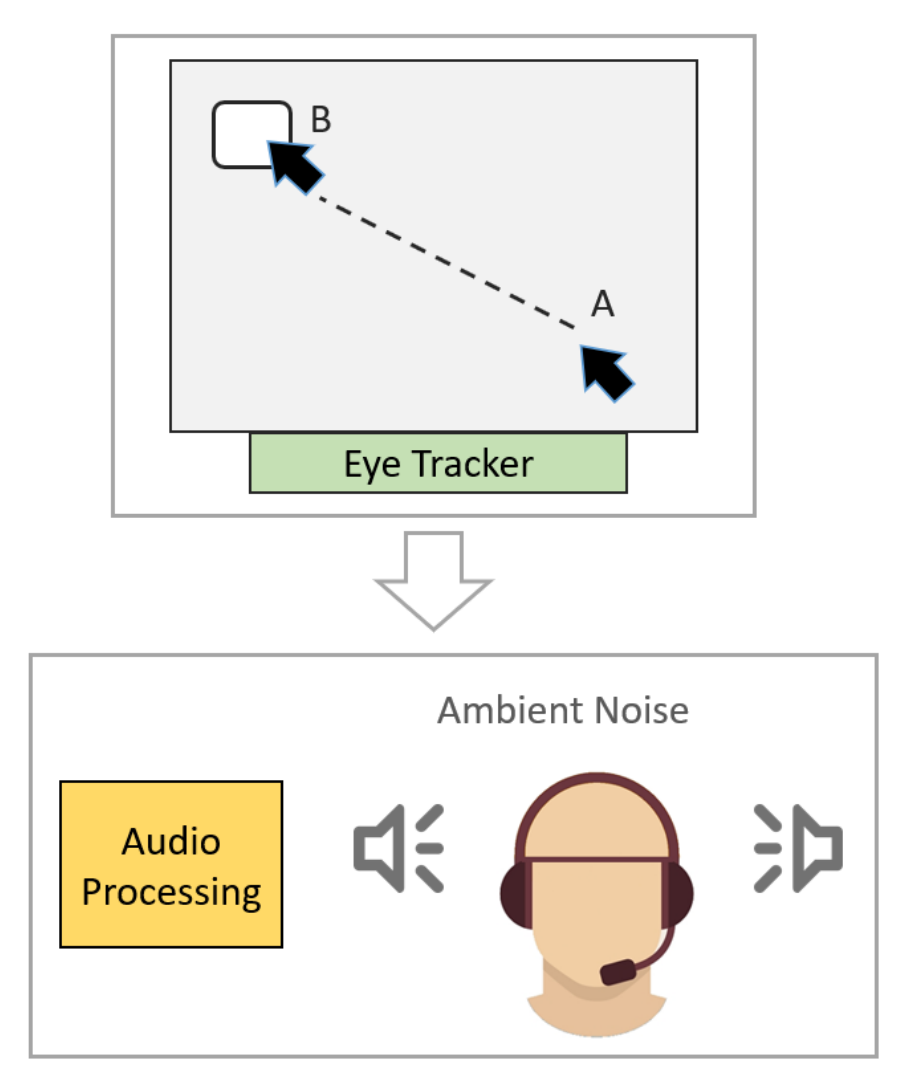}
	\caption{EyeTAP system: The eye tracker is used to move the pointer from A to B. The user makes an acoustic pulse by mouth and the signal processing module interprets the signal as an input and triggers a click event to select B. The system has an ambient noise tolerance of up to 70 dB.}
	\label{fig:system_overview}
\end{figure}

A simple mouse interaction consists of moving the pointer to a target (pointing phase), and clicking on it to trigger a function (selection phase). In the EyeTAP prototype the mouse pointer position is captured using the Tobii 4C tracker ~\footnote{https://tobiigaming.com/product/tobii-eye-tracker-4c/} and selection is done by generating an acoustic pulse by mouth (e.g. a mouth click) which is captured by a headset microphone (Logitech H370). The EyeTAP prototype was developed and the experiments were run on a commodity computer system: 64-bit Windows 10 PC with Intel i7 2.67GHz CPU, 12 GB RAM, 1 TB hard disk and NVIDIA GeForce GTX 770 graphics card. Thus, EyeTAP is a cost-effective system that can be applied at almost any work space. Figure \ref{fig:system_overview} illustrates the EyeTAP system setup.

\subsection{Eye Tracking: Pointing Phase}
The Tobii SDK (TobiiEyeXSdk$-$Cpp$-$1.8.498) supports different events related to eye tracking activities such as providing the location of the current eye gaze, positions of both eyes, fixation points and user presence in front of the eye tracker. We employed the eye gaze library (API) to obtain users' gaze locations. These locations show the current gaze position on the screen as pixels. The SDK supports eye movements in a 3D coordinate system (horizontal, vertical, depth) but we applied a 2D coordinate system (x,y) such that the mouse cursor was synchronized with the gaze positions to control the mouse pointer on the screen. Eye-tracking for the EyeTAP prototype was developed in C++ and integrated as a new plug-in into the Tobii SDK.

\subsection{Auditory Processing: Selection Phase}
To simulate a click on the item to be selected a headset microphone listens to the user while suppressing the background ambient sounds/noise (conversations in office and equipment sounds) in real-time. The intensity of the mouth noise and distance of microphone is adjusted by the user before the test. A detected pulse in the real-time audio signal (a value larger than a predefined threshold) is regarded as a click. The threshold's value can be adjusted based on the environment to reduce background ambient noise. The EyeTAP prototype has an ambient noise tolerance of up to 70 dB. When a significant increase in the frequency spectrum (greater than the threshold) is detected a mouse click event is triggered. In general, recording is categorized into two phases: audible and silent periods. Any audible period with an intensity greater than the predefined threshold will be detected as an input signal to the system as the binary 1; similarly, values smaller than the threshold value are regarded as binary 0. The intuition behind the auditory processing was inspired from the simplicity of the Morse code~\cite{morse_code}, which consists of a series of ON/OFF signals triggered by tone or light. Information is interpreted using dots and dashes and therefore can be used to represent transmitted signals through a sequence of True/False variables. 

\section{Evaluation} 
\label{Evaluation}
To evaluate the effectiveness of the developed EyeTAP prototype, we ran two independent user studies each with two internal experiments with 33 participants (13 female, from 22 to 35 years old, SD=2.96). All subjects partook in both experiments. Prior to running the experiments, subjects were informed about the purpose of the study, trained on each of the methods to be tested, and participated in a pre-test questionnaire probing them on their background in the fields of eye tracking, voice recognition technologies and their preferred kind of interaction in the case of hands-free alternatives. The Tobii calibration software was used to calibrate the system for each participant before starting the study. At the end of the two experiments subjects filled out a post-test questionnaire, which consisted of the NASA TLX questionnaire \cite{nasa_tlx} followed by specific questions about the subjects' perceptions of the different interaction methods. The order of interaction method was randomly selected for each participant. We played an artificial ambient noise through stereo desktop speakers of 50 dB to simulate a typical work environment since EyeTAP and voice recognition rely on audio inputs.

\subsection{User Study 1: Matrix-based Test}
In the first experiment, the EyeTAP interaction method was compared with: (a) the mouse, (b) dwell-time, and (c) eye tracking with voice-recognition. In this experiment, a matrix of buttons (targets), were randomly distributed across the screen. The task of the subjects was to point and click on buttons shown on the screen in increasing numerical order for various levels of difficulty from 1 (easy) to 5 (hard), described in detail below. The order of interaction methods seen by each subject was randomly selected for each participant however, the level of difficultly was presented in ascending order. 

\subsubsection{Stimulus} 
\begin{figure}[htp]
	\centering
		\includegraphics[scale=0.35]{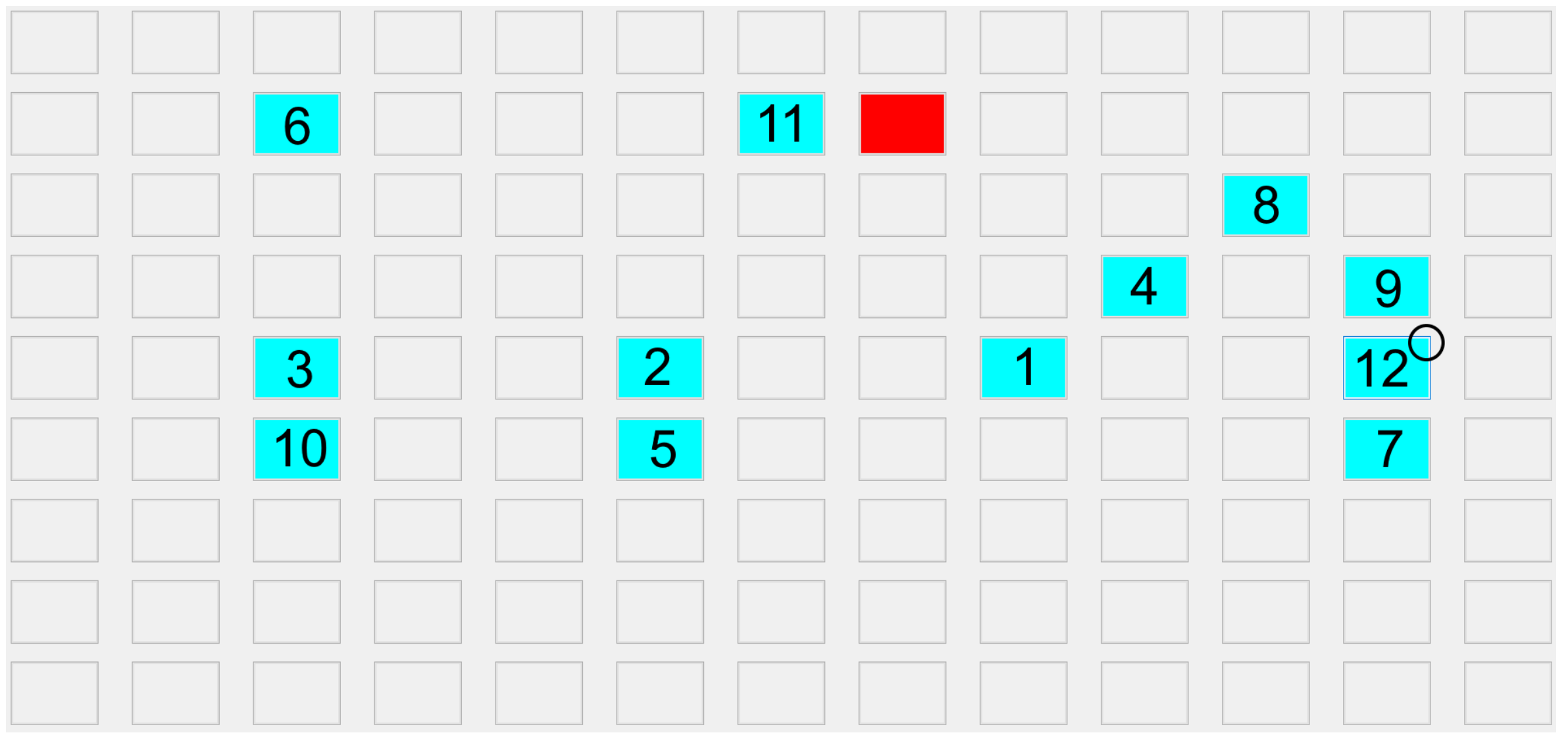}
	\caption{The matrix-based test for difficulty level 5. Target buttons are distributed randomly across the screen. The red button illustrates an error. The black circle on number 12 shows the current eye gaze location. Labels were enlarged for higher visibility.}
	\label{fig:test_screenshot}
\end{figure}

The stimulus consisted of 77 buttons (11 columns $\times$ 7 rows) some labeled with numbers and others not, which covered the entire screen at a resolution of 1920 $\times$ 1080 pixels on a Dell P2411Hb monitor. Two marginal columns (far left, far right) and two rows (top, bottom) were removed from the active selection due to the high difficulty to be selected by users during the pilot-test. Buttons that were not labeled are considered as \textit{barriers} or \textit{distractions}. To provide feedback to the subject, labeled buttons change color after the user has successfully pointed and selected on the correct button. Wrongly selected barriers (buttons with no label) are highlighted in red. The level of difficulty of the stimulus was also increased across subject trials. This was done by increasing the number of targets that had to be selected by the subject. Five levels of difficulty were used for each interaction method: level 1 (4 targets), level 2 (6 targets), level 3 (8 targets), level 4 (10 targets) and level 5 (12 targets). Targets were randomly distributed over the entire screen for each level. Figure \ref{fig:test_screenshot} shows the matrix-based test during difficulty level 5. The cursor that was used was a black circle because it is easier for users to keep it on the target's boundary rather than a pointer.

\subsubsection{Mouse}
For the mouse method (our baseline method for comparison), subjects simply used a mouse to move to targets and select them in numerical order.

\subsubsection{Dwell-time}
For the dwell-time method, where an internal timer is used to determine if a target was selected. The range of dwell-time is in (300-1100) milliseconds for target selection \cite{vspakov2004line}. Then we defined the target activation threshold to 500 milliseconds, since it showed best performance in \cite{mackenzie2012evaluating} and participants preferred a dwell-time around 500 ms in a user study \cite{vspakov2004line}. In other words, a target was selected when a subject focused on a target for 0.5 seconds, and if the subject moved their gaze away from the target prior to 0.5 seconds the target selection process would restart. 

\subsubsection{Eye Tracking with Voice recognition}
For voice recognition, eye tracking was used for pointing and voice for selection. The method was developed using the built-in Windows 10 speech recognition capabilities available in the .NET framework. We implemented a C\# application to respond to the activation keyword 'select' to trigger a mouse click. The same microphone was used as for the EyeTAP test.

\subsubsection{Measures}
\label{MatrixBasedVariablesDetails}
The following variables were recorded: \textit{completion time}, \textit{path cost of selecting targets}, \textit{error locations}, and \textit{cognitive load} (based on the NASA TLX scores). An internal logging module recorded subjects' actions, selection times, as well as the number of correct and wrong selections.  For the path cost measure the shortest path between targets and the produced path by each interaction method was processed. To compare the shapes of the generated paths, we used the dynamic time warping (DTW) algorithm~\cite{1104847, 1163491,1163055}. Since DTW works on a time-value domain the paths produced by the eye tracker were decomposed into their horizontal and vertical values and compared with their associated shortest path models' \textit{X} and \textit{Y} values. We applied the built-in \textit{DTW} function in the Python DTW 1.3.3 module \footnote{https://github.com/pierre-rouanet/dtw} to measure the deviations of each path from the shortest path model.

\subsection{User Study 1: Dart-based Test}
The purpose of this experiment was to measure the accuracy of EyeTAP in comparison to the previously proposed eye-based interaction methods. The task of the subject was to select, as accurately as possible, the bull's-eye of a dart target using each interaction method. In this experiment, the eye tracker was used for the pointing phase for each of the interaction methods, however selection of the target was triggered by different methods, i.e. dwell-time, voice command or EyeTAP acoustic signal. In order to take into consideration the fact that eye tracking has different accuracy in different regions of the monitor, we computed an average value based on five trials for each interaction method where the stimulus was shown at different areas of the screen near the center of the screen randomly. Each new randomly chosen trial began two seconds after selection of the previous target, allowing users time to change their gaze and to focus on the new target. For the dwell-time method, a countdown (5 to 0) representing remaining 100 milliseconds was displayed during the selection phase and users needed to focus on the dart shape before this time was up.

\subsubsection{Stimulus}
\begin{figure}[htbp]
	\centering
		\includegraphics[width=0.23\textwidth]{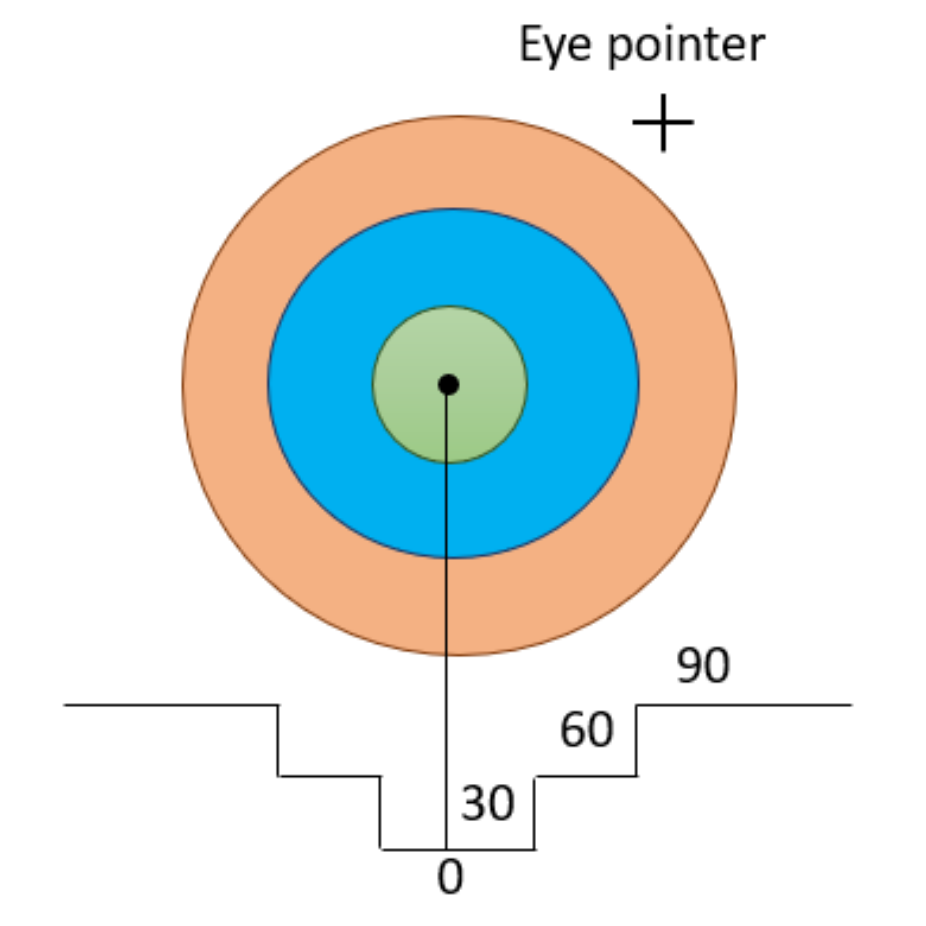}
	\caption{Dart-based test stimuli: the accuracy is highest in the green area. The cross-hair icon indicates the correct eye gaze location.}
	\label{fig:circular_based}
\end{figure}

The stimulus for this experiment consisted of a dart-like target with three circles, green (0 to 30 pixels radius), blue (30 to 60 pixels radius) and red (60 to 90 pixels radius) as in Figure \ref{fig:circular_based}. Points within the center area i.e. green have the lowest range of distances to the bulls-eye; each other co-centric circle has a larger range of distance values. Any point lying outside the three co-centric circular areas is considered as having a fixed maximum distance of 90 pixels. For this experiment, a cross-hair icon was used.

\subsubsection{Measures}
The purpose of this test was to measure the selected point's distance on the dart target to the center of the core circle (in green), thus the accuracy is measured in pixels. Since the measured trials are chosen randomly, the average is calculated to compare different methods based on accurate selection. 

\subsection{User Study 2: Ribbon-shaped Test}
In order to compare our method to other studies, we performed the FittsStudy~\cite{wobbrock2011effects}. This study is used to analyze pointing interaction methods in accordance to well-established academic standards. As part of this study, we measured three metrics to compare the performance of all interaction techniques for point-and-select tasks, (1) \textit{throughput}, (2) \textit{movement time} and (3) \textit{error rates} for ribbon-shaped targets (see figure \ref{fig:fitts_overview}). We applied the FittsStudy application \footnote{http://depts.washington.edu/acelab/proj/fittsstudy/index.html} by Wobbrock \emph{et al.} \cite{wobbrock2011effects}. The test session includes three distances (256, 384, 512) and two widths (96, 128) pixels.

\subsection{User Study 2: Circle-shaped Test}
\begin{figure}[htbp]
	\centering
		\includegraphics[width=0.25 \textwidth]{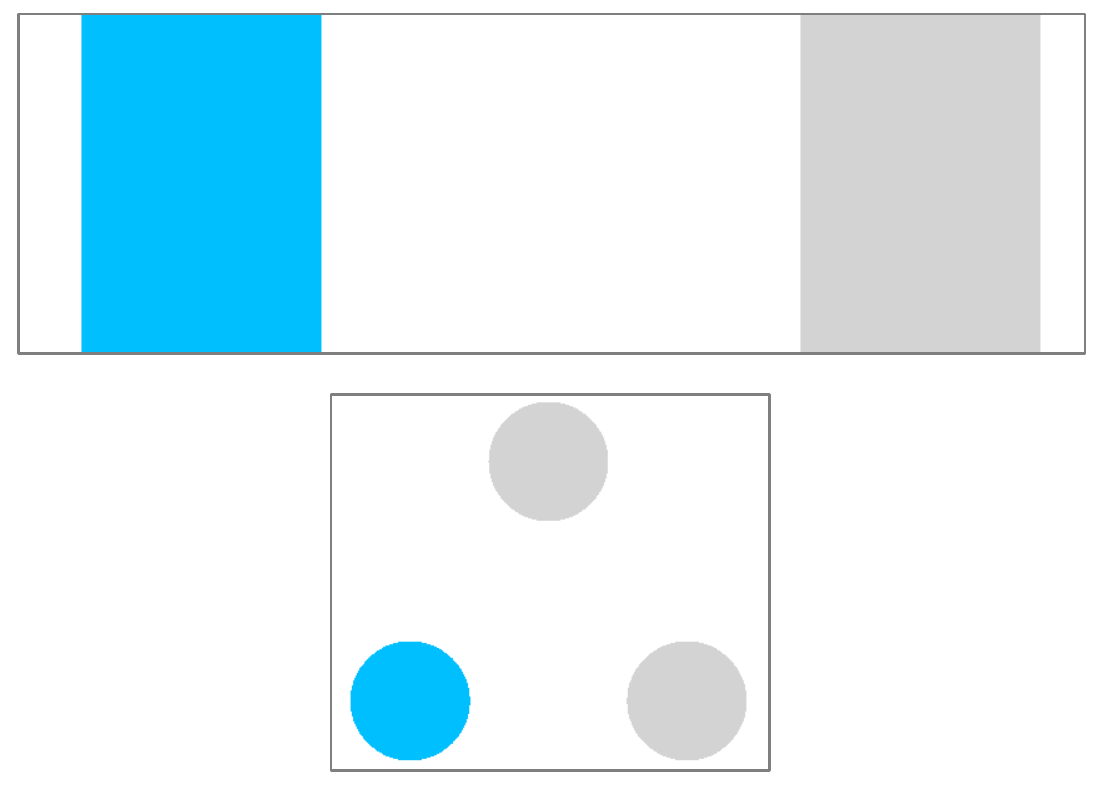}
	\caption{Screenshots of the 'FittsStudy' application \cite{wobbrock2011effects}. Top figure illustrates the ribbon-shaped stimuli and the bottom figure shows the circle-shaped stimuli. The highlighted targets are shown in blue to represent the active target to be selected.}
	\label{fig:fitts_overview}
\end{figure}

This test is similar to the Ribbon-shapped test, however, contains different target shapes. Figure \ref{fig:fitts_overview} illustrates the screenshots of both test applications. This experiment contains uni-variate endpoint deviation (\textit{SD\textsubscript{x}}) through one axis and bi-variate endpoint deviation (\textit{SD\textsubscript{x,y}}) through both axes for throughput calculations which results in better Fitts' law model \cite{wobbrock2011effects}.

\section{Results}

To determine the effectiveness of the EyeTAP method, we analyzed the results of our experiments using an analysis of variance (ANOVA) followed by Bonferroni posthoc tests with the IBM SPSS software \footnote{https://www.ibm.com/analytics/spss-statistics-software}.

\subsection{User Study 1: Matrix-based User Study}
A two-way repeated measures ANOVA (methods $\times$ difficulty levels) was performed to examine the effect of interaction type on: (1)  \textit{completion time} and (2) \textit{path costs of target selection} for each method and difficulty levels. 	

\subsubsection{Completion time}
\begin{figure}[htbp]
	\centering
		\includegraphics[width=0.47 \textwidth]{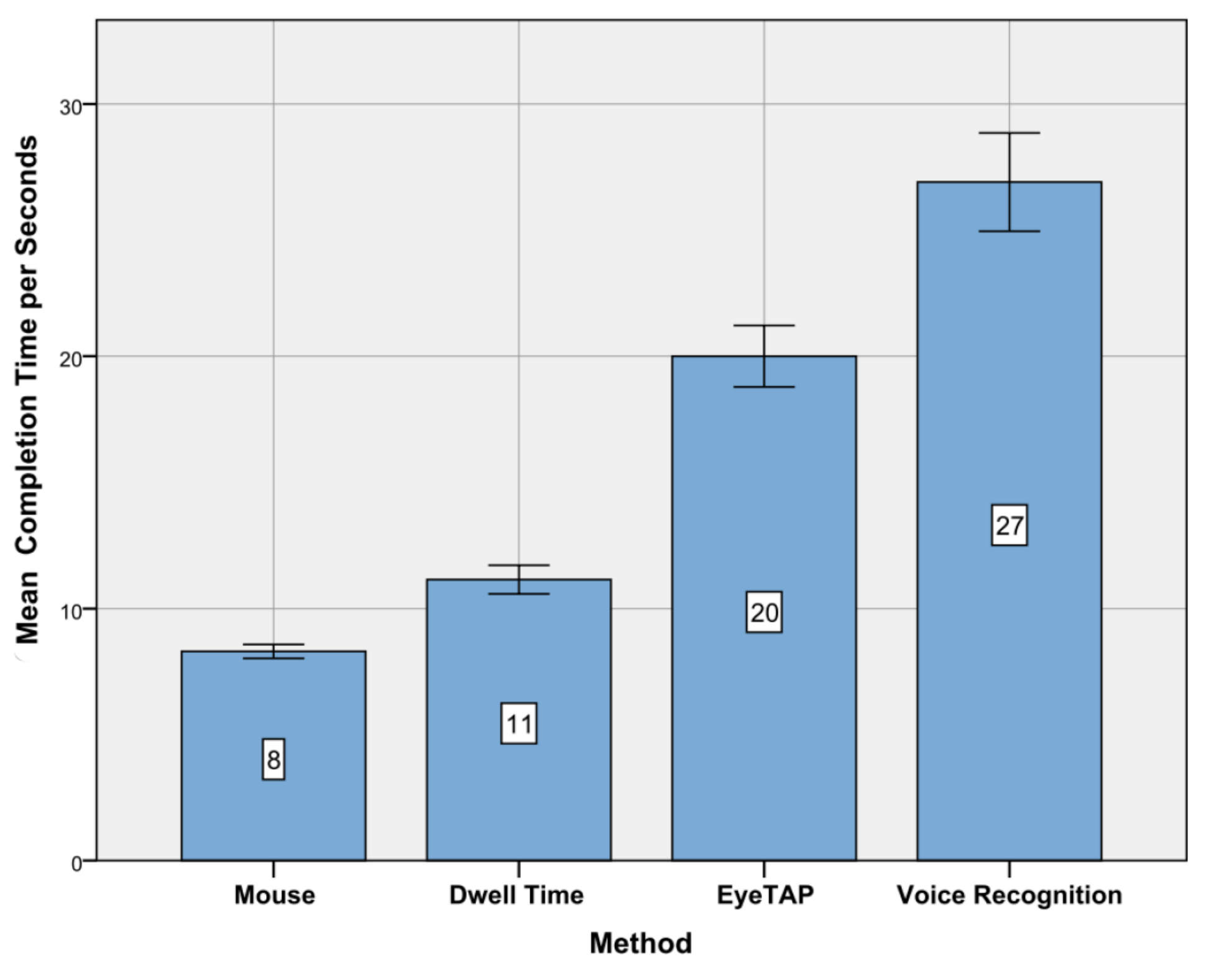}
	\caption{Average completion time of point-and-select tasks for all participants obtained from the matrix-based user study for 8 targets per level ($\frac{40~targets}{5~levels}$). Completion time was significantly different for all techniques ($p < .001$).}
	\label{fig:matrix_completion_time}
\end{figure}

We found a significant effect of interaction method on completion time (F(12,384)=8.51, \textit{p} < .001). A posthoc Bonferroni comparison test showed a significant difference between mouse ($M=8017.955~ms$, $SE=645.433~ms$) and all other eye tracking methods (see figure~\ref{fig:matrix_completion_time}). In addition, EyeTAP ($M=19998.812~ms$, $SE=2122.329~ms$), dwell-time ($M=11154.830~ms$, $SE=788.395~ms$) and voice recognition ($M=26904.333~ms$, $SE=2467.576~ms$) are significantly different (\textit{p} < .05). Figure \ref{fig:matrix_completion_time} illustrates the average completion time per method for 8 targets per level ($\frac{40~targets}{5~levels}$).



\subsubsection{Path costs of target selections}
\begin{figure}[htbp]
		\includegraphics[width=0.47 \textwidth]{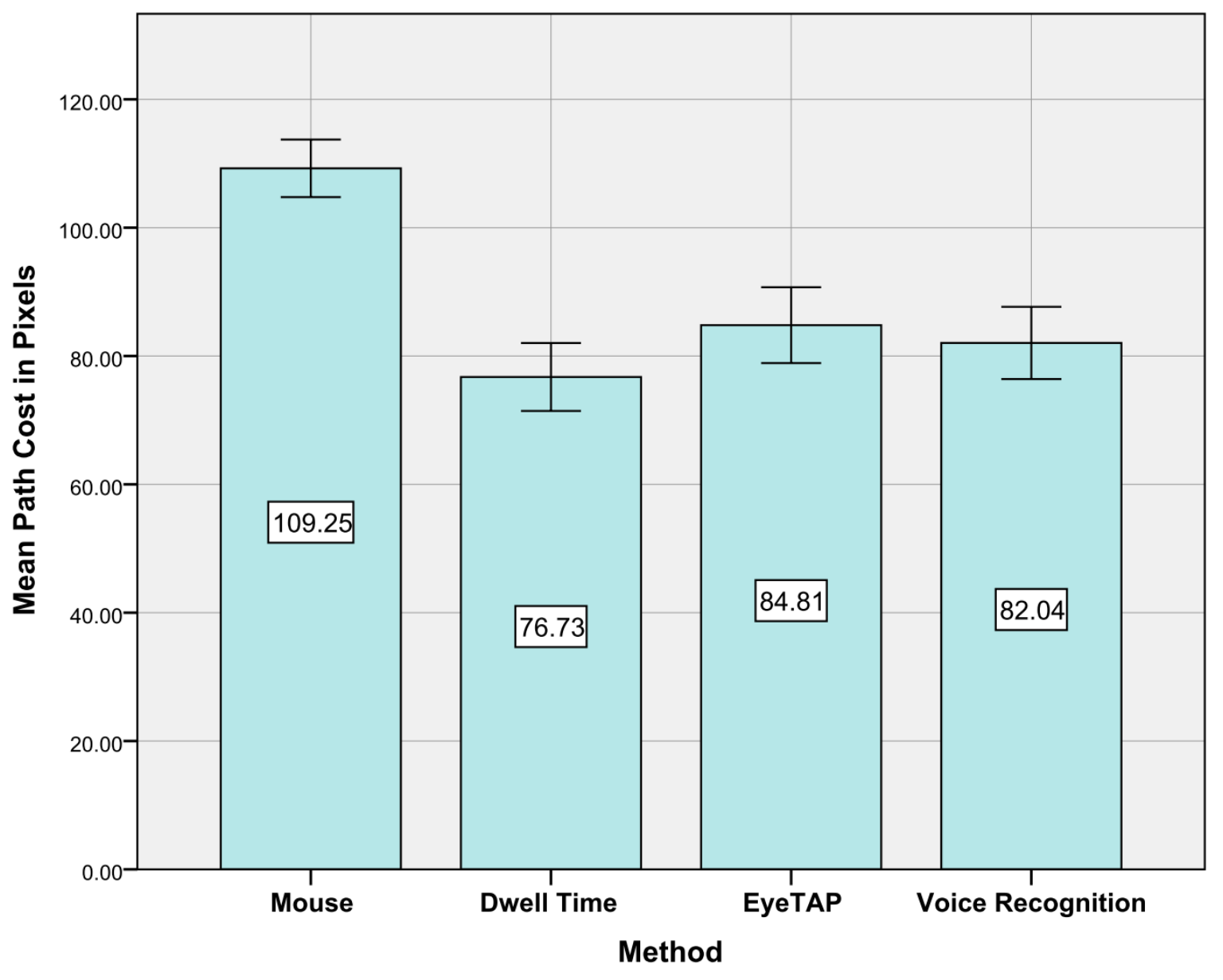}
	\caption{Mean path cost comparison calculated using the dynamic time warping (DTW) algorithm. All eye tracking techniques have shorter path lengths than mouse interaction for traversing items on a screen ($p < .05$).}
	\label{fig:dtw_all}
\end{figure}

To examine the paths produced by selecting targets we compared the original locations of the targets and the shortest path (ideal path model), as described in Section \ref{Evaluation}. For each method, we had a $\frac{distance}{cost}$ measure to the shortest path. This metric can be regarded as the \textit{footprint} of each interaction technique on the display. A two-way repeated measures ANOVA (methods $\times$ difficulty levels) showed that there was a significant effect of interaction type on path cost (F(12,384)=2.57, \textit{p} < .05). A Bonferroni posthoc test showed that dwell-time ($M=76.73~pixels$, $SE=5.09~pixels$) produced the shortest path among all other interaction techniques, even better than the mouse interaction ($M=109.25~pixels, SE=3.82~pixels$)
with \textit{p} < .05. However, there is no significant difference between dwell-time ($M=76.73~pixels$, $SE=5.09~pixels$), EyeTAP ($M=84.80~pixels$, $SE=3.59~pixels$) and voice recognition ($M=82.03~pixels$, $SE=4.41~pixels$). Figure \ref{fig:dtw_all}, which shows the path costs for all interaction methods, reveals that eye tracking movements produce significantly lower movements than mouse on a large screen. 


\subsubsection{Errors in target selections}	
\begin{figure}[htbp]
	\centering
		\includegraphics[width=0.43 \textwidth]{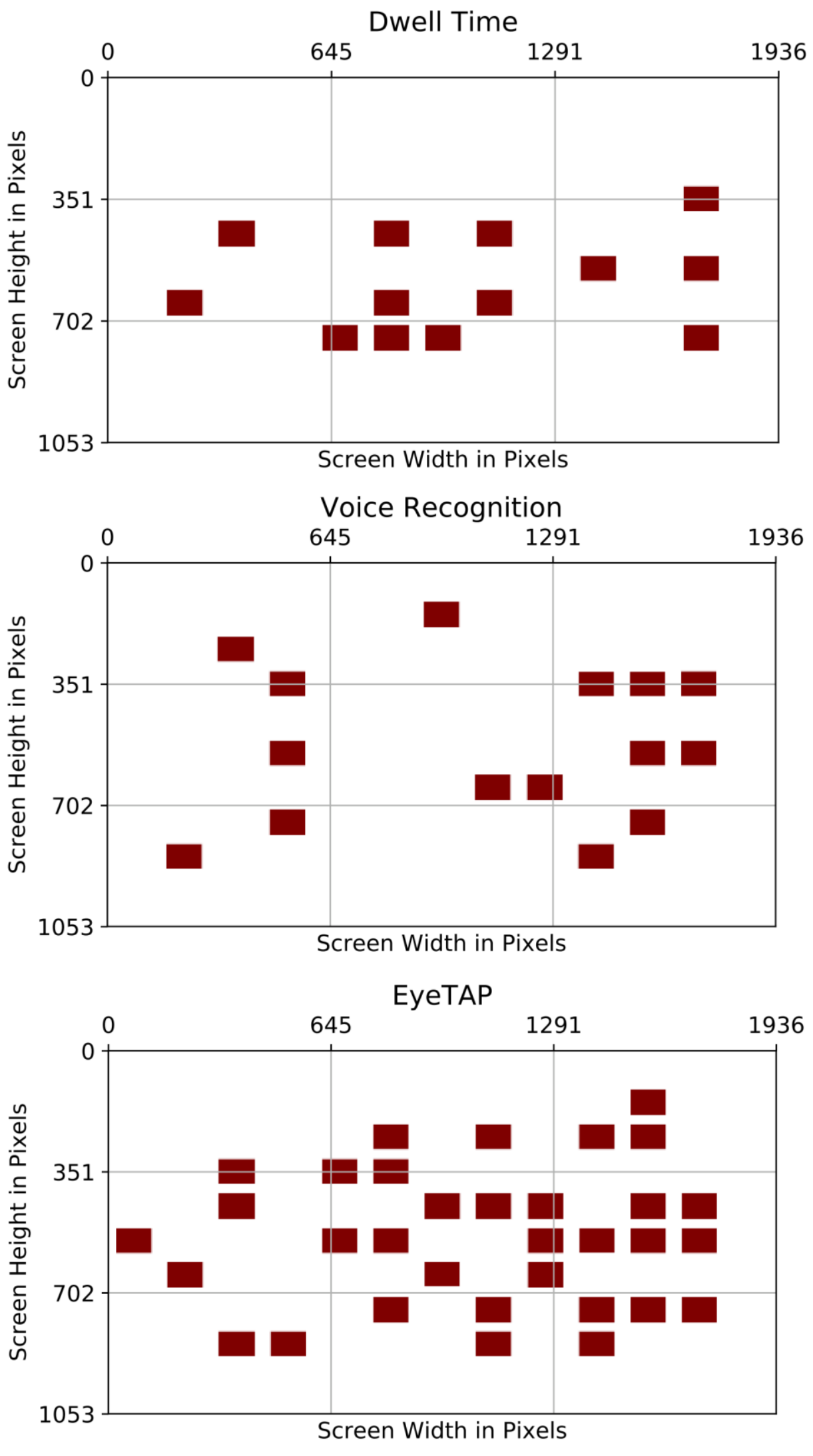}
	\caption{The locations of errors during the matrix-based user study (figure \ref{fig:test_screenshot}) for difficulty level 5. The right side of the screen as well as bottom side are more error prone than the left and top sides.}
	\label{fig:error_locations}
\end{figure}

To measure the effectiveness of each Midas touch solution we need to consider a penalty for wrongly selected neighboring targets. Those targets are shown in red on the screen (see figure \ref{fig:test_screenshot}). We projected the locations of errors per each interaction method, since difficulty level 5 has the highest number of targets (12 targets) on the screen, we illustrate the locations for this difficulty level in Figure \ref{fig:error_locations}. EyeTAP has the highest number of errors, however the figure reveals the potential regions of the screen which are more error prone. As shown in the figure, most errors occurred from the center towards the right side of the screen. In fact, the right side of the screen produces more errors than the left side. Moreover, the lower side produces more errors than the top side. Feit \emph{et al.} showed that the same bottom and right regions of the screen have lower accuracy \cite{feit2017toward}. We confirm their results and also demonstrate that the same regions are also more error prone.

\subsection{User Study 2: Dart-based User Study} 

\begin{figure}[htbp]
	\centering
		\includegraphics[width=0.47 \textwidth]{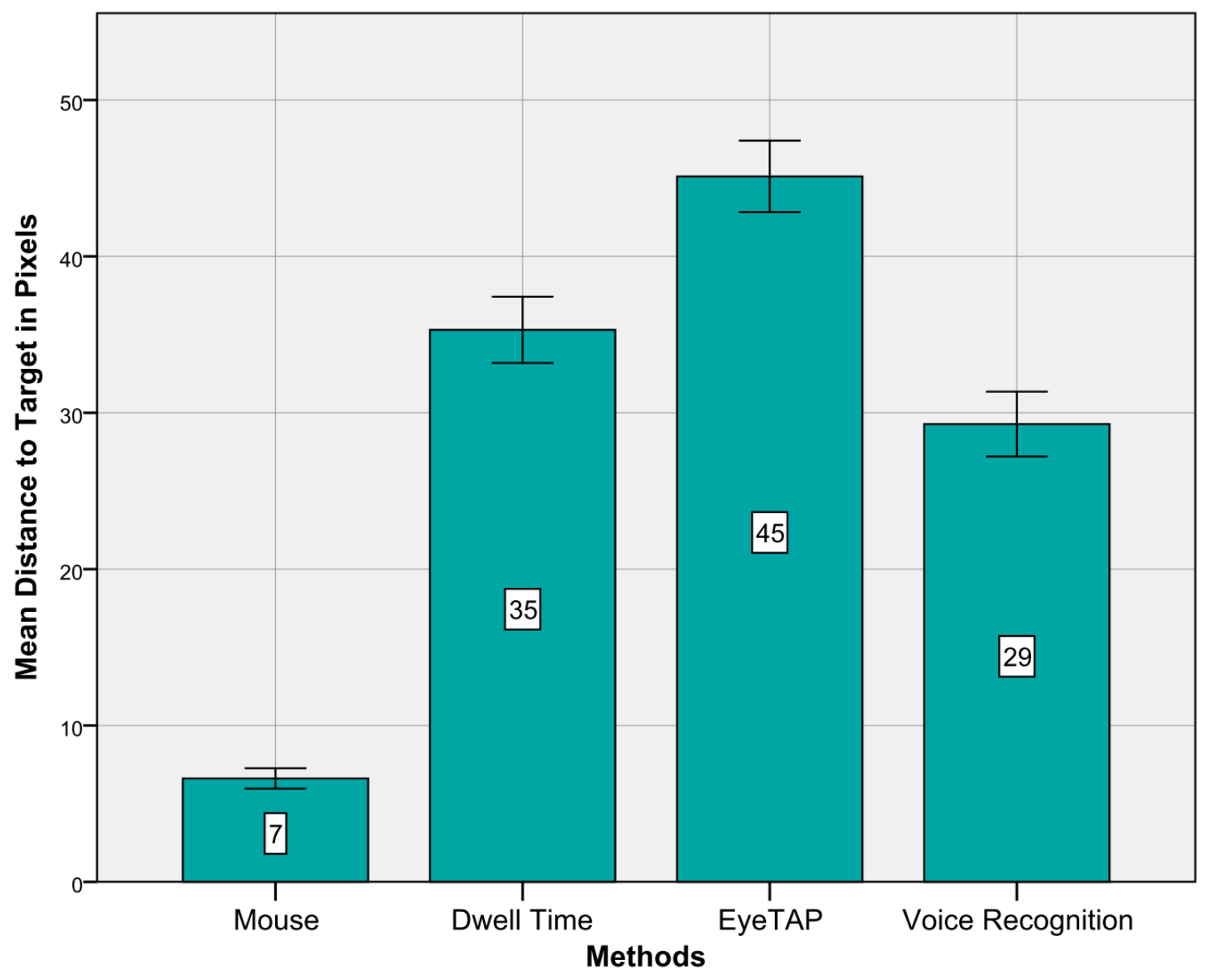}
	\caption{The mean distance to target in pixels for dart-based experiment ($p < .001$).}
	\label{fig:dart_results_chart}
\end{figure}

We performed a one-way repeated measures ANOVA to compare the effect of the different interaction methods on accuracy. The results of the ANOVA showed all eye tracking methods have statistical difference (F(3,96)=104.92, \textit{p} < 0.001) on selection accuracy. In fact, the mouse interaction has the lowest distance to target (higher accuracy) compared to eye tracking techniques. EyeTAP ($M=45.11~pixels$, $SE=2.28~pixels$) achieved the highest mean pixel accuracy compared to dwell-time ($M=35.30~pixels$, $SE=2.11~pixels$) and voice recognition ($M=29.27~pixels$, $SE=2.07~pixels$). Figure \ref{fig:dart_results_chart} depicts the results of the accuracy test.

\subsection{User Study 2: Ribbon-shaped Test}
A one-way repeated measures ANOVA was performed to examine the effect of interaction type on: (1)  \textit{movement time}, (2) \textit{throughput} and (3) \textit{error rates} for each interaction method.

\subsubsection{Movement time}
\begin{figure}[htbp]
		\includegraphics[width=0.47 \textwidth]{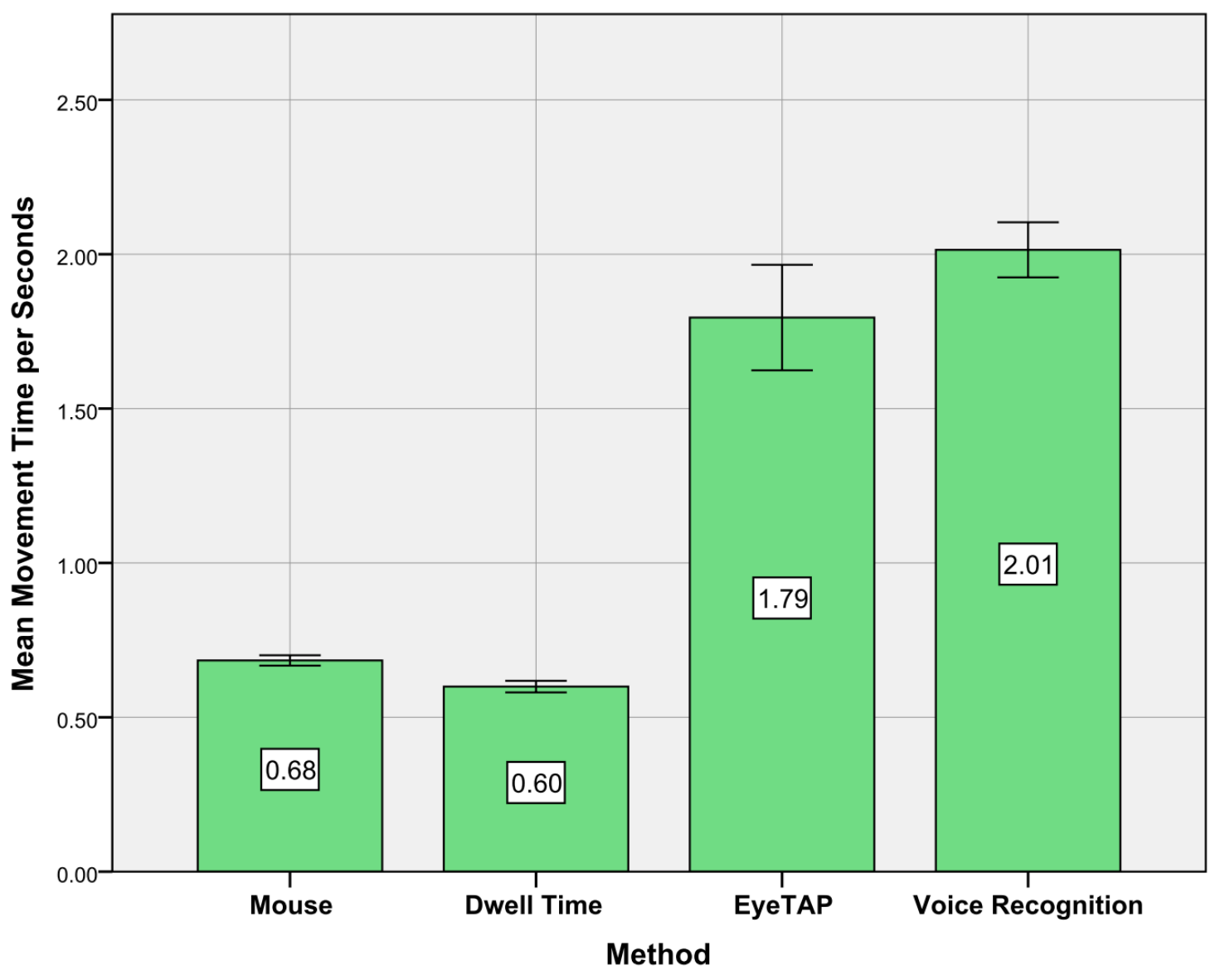}
	\caption{The calculated movement time per method for the ribbon-shaped test ($p < .001$).}
	\label{fig:ribbon_movement_time}
\end{figure}

We found a significant effect of the interaction method on movement time (F(3,96)=69.42, \textit{p} < .001). A posthoc Bonferroni comparison test showed a significant difference between mouse ($M=684.15~ms$, $SE=16.80~ms$) and all other eye tracking methods (figure~\ref{fig:ribbon_movement_time}). In addition, among all eye tracking methods, dwell-time ($M=599.39~ms$, $SE=18.76~ms$) achieved significantly lower movement time than EyeTAP ($M=1794.89~ms$, $SE=170.90~ms$) and voice recognition ($M=2014.20~ms$, $SE=89.28~ms$) techniques. However, there is no statistical significance between EyeTAP and voice recognition. The lower movement time of dwell-time method compared to the mouse interaction is associated with the low activation time (500 ms).

\subsubsection{Throughput}
\begin{figure}[htbp]
	\centering
		\includegraphics[width=0.47 \textwidth]{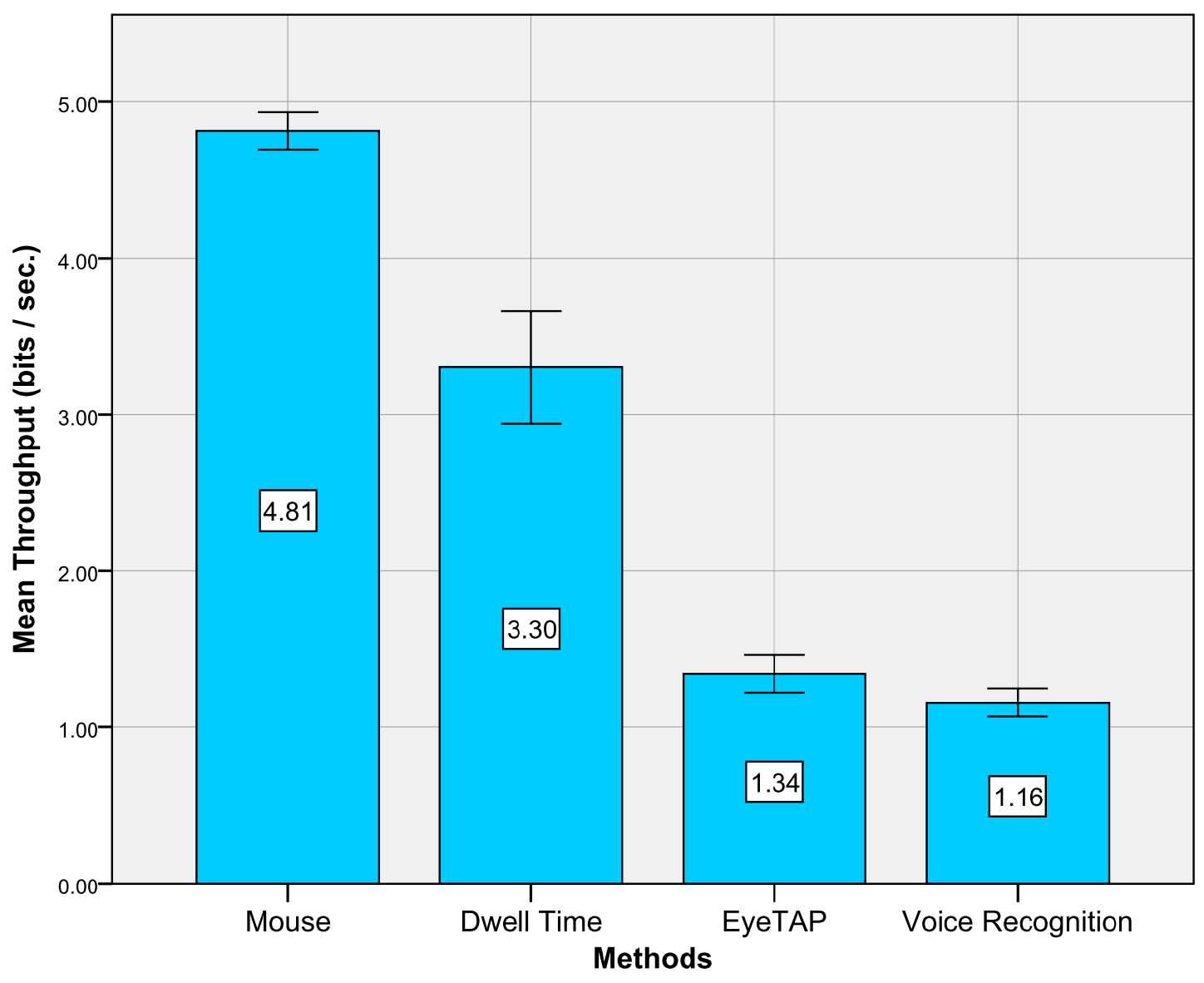}
	\caption{The calculated throughput per method for the ribbon-shaped test ($p < .001$).}
	\label{fig:ribbon_tp}
\end{figure}

We found a significant effect of the interaction method on throughput (F(3,96)=75.13, \textit{p} < .001). A posthoc Bonferroni comparison test showed a significant difference between dwell-time ($M=3.30~bits/sec$, $SE=0.36~bits/sec$) and all eye tracking methods (figure~\ref{fig:ribbon_tp}). The mouse ($M=4.81~bits/sec$, $SE=0.11~bits/sec$) achieved higher throughput than the eye tracking methods. However, there is no statistical difference between voice recognition ($M=1.15~bits/sec$, $SE=0.09~bits/sec$) and EyeTAP ($M=1.34~bits/sec$, $SE=0.12~bits/sec$).

\subsubsection{Error rates}
\begin{figure}[htbp]
	\centering
		\includegraphics[width=0.47 \textwidth]{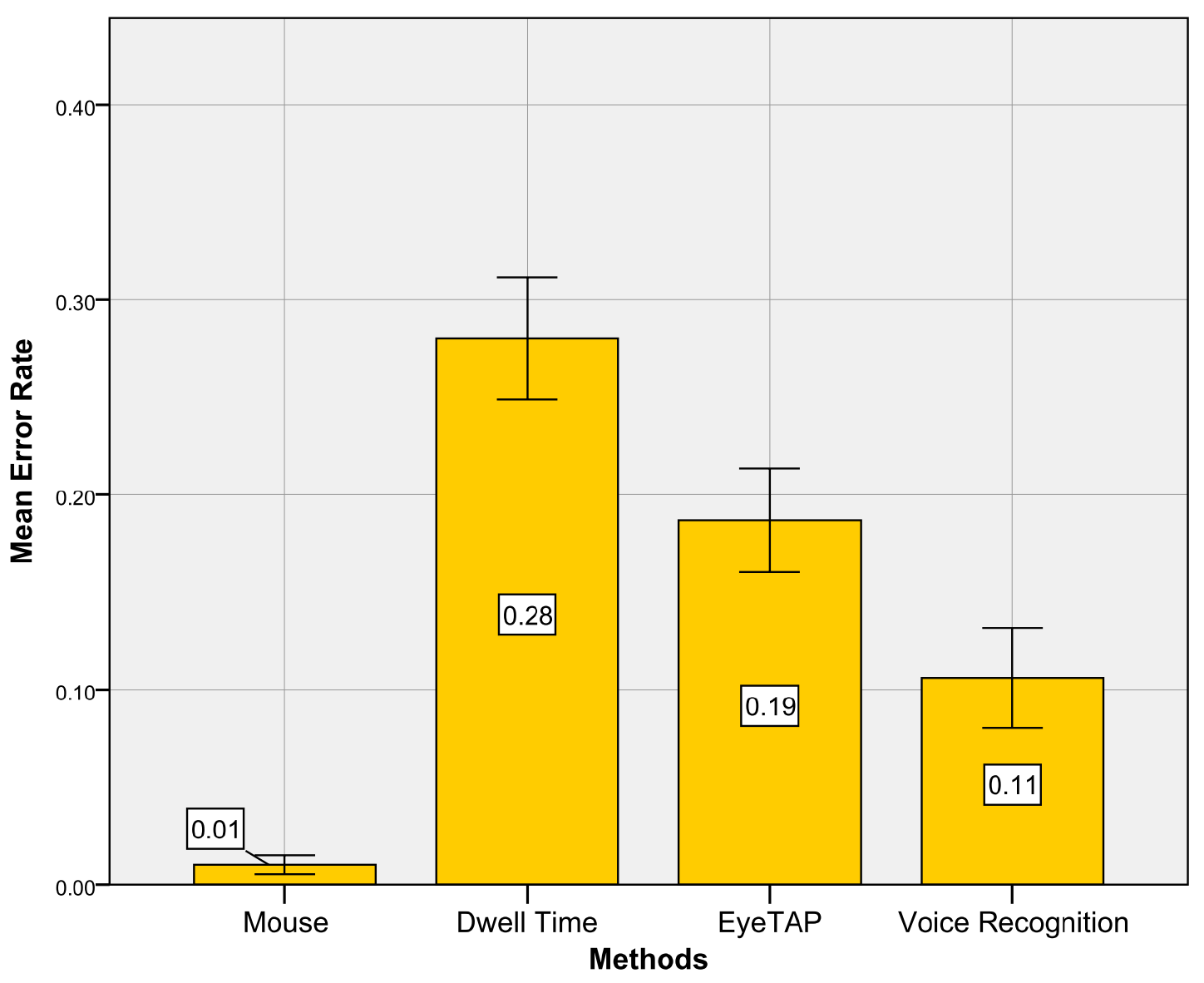}
	\caption{The calculated error rates per method for the ribbon-shaped test ($p < .001$).}
	\label{fig:ribbon_error_rates}
\end{figure}

We found a significant effect of the interaction method on error rates (F(3,96)=27.15, \textit{p} < .001). A posthoc Bonferroni comparison test showed a significant difference between mouse ($M=0.01~errors$, $SE=0.005~errors$) and all eye tracking interactions (see Figure~\ref{fig:ribbon_error_rates}). In addition, dwell-time ($M=0.28~errors$, $SE=0.03~errors$) reached a higher error rate than EyeTAP ($M=0.18~errors$, $SE=0.02~errors$) and voice recognition ($M=0.10~errors$, $SE=0.02~errors$).

\subsection{User Study 2: Circle-shaped Test}
A one-way repeated measures ANOVA was performed to examine the effect of interaction type on: (1)  \textit{movement time}, (2) \textit{throughput} and (3) \textit{error rates} for each interaction method. This experiment is similar to ribbon-shaped test but contains an extra metric to measure throughput of each method.

\subsubsection{Movement time}
\begin{figure}[htbp]
	\centering
		\includegraphics[width=0.47 \textwidth]{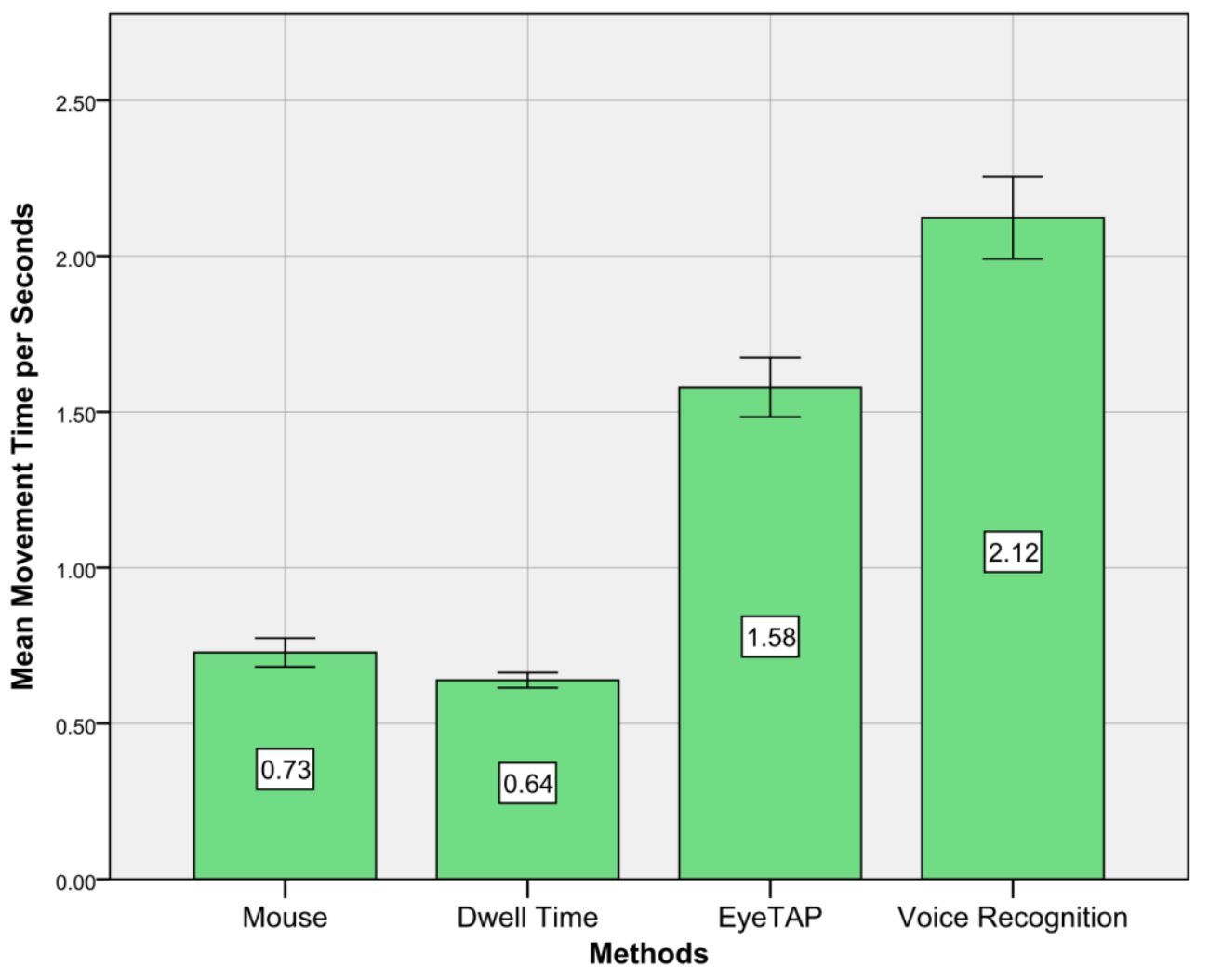}
	\caption{The calculated movement time per method for the circle-shaped test ($p < .001$).}
	\label{fig:circle_movement_time}
\end{figure}

We found a significant effect of the interaction method on movement time (F(3,96)=67.48, \textit{p} < .001). A posthoc Bonferroni comparison test showed a significant difference between EyeTAP ($M=1578.95~ms$, $SE=95.34~ms$), dwell-time ($M=638.80~ms$, $SE=24.35~ms$), voice recognition ($M=2123.35~ms$, $SE=132.42~ms$) and mouse ($M=727.91~ms$, $SE=46.12~ms$). However, there is no statistical difference between mouse ($M=727.91~ms$, $SE=46.12~ms$) and dwell-time ($M=638.80~ms$, $SE=24.35~ms$). Figure \ref{fig:circle_movement_time} illustrates the mean movement time per method for the circle-shaped test.

\subsubsection{Throughput}
\begin{figure*}[htbp]
	\centering
		\includegraphics[width=0.73 \textwidth]{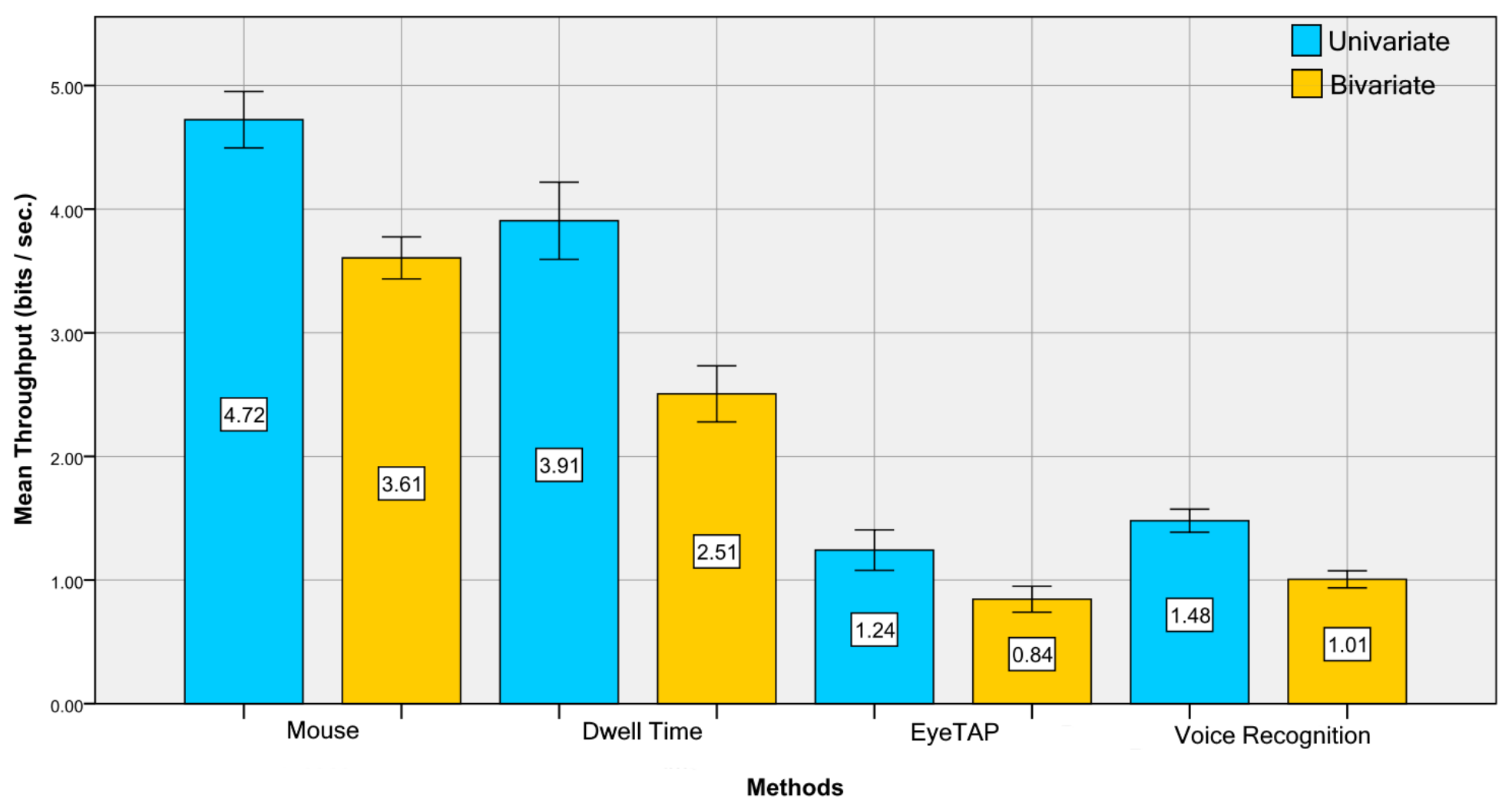}
	\caption{The calculated throughput for both uni-, and bi-variations per method for the circle-shaped test ($p < .001$).}
	\label{fig:circle_tp_chart}
\end{figure*}

Since the circle-shaped test contains two variations (uni-variate, bi-variate) to measure throughput \cite{wobbrock2011effects}, we ran a two-way repeated measures ANOVA (throughput $\times$ variation) and found a significant effect of the interaction method on throughput (F(3,96)=19.75, \textit{p} < .001). A posthoc Bonferroni comparison test showed a significant difference between
mouse ($M=4.16~bits/sec$, $SE=0.18~bits/sec$), dwell-time ($M=3.20~bits/sec$, $SE=0.25~bits/sec$), voice-recognition ($M=1.24~bits/sec$, $SE=0.07~bits/sec$) and EyeTAP ($M=1.04~bits/sec$, $SE=0.13~bits/sec$). However, there is no statistical difference between voice-recognition ($M=1.24~bits/sec$, $SE=0.07~bits/sec$) and EyeTAP ($M=1.04$\newline$bits/sec$, $SE=0.13~bits/sec$). Figure \ref{fig:circle_tp_chart} shows both variations of throughput per interaction method.

\subsubsection{Error rates}
\begin{figure}[htbp]
	\centering
		\includegraphics[width=0.47 \textwidth]{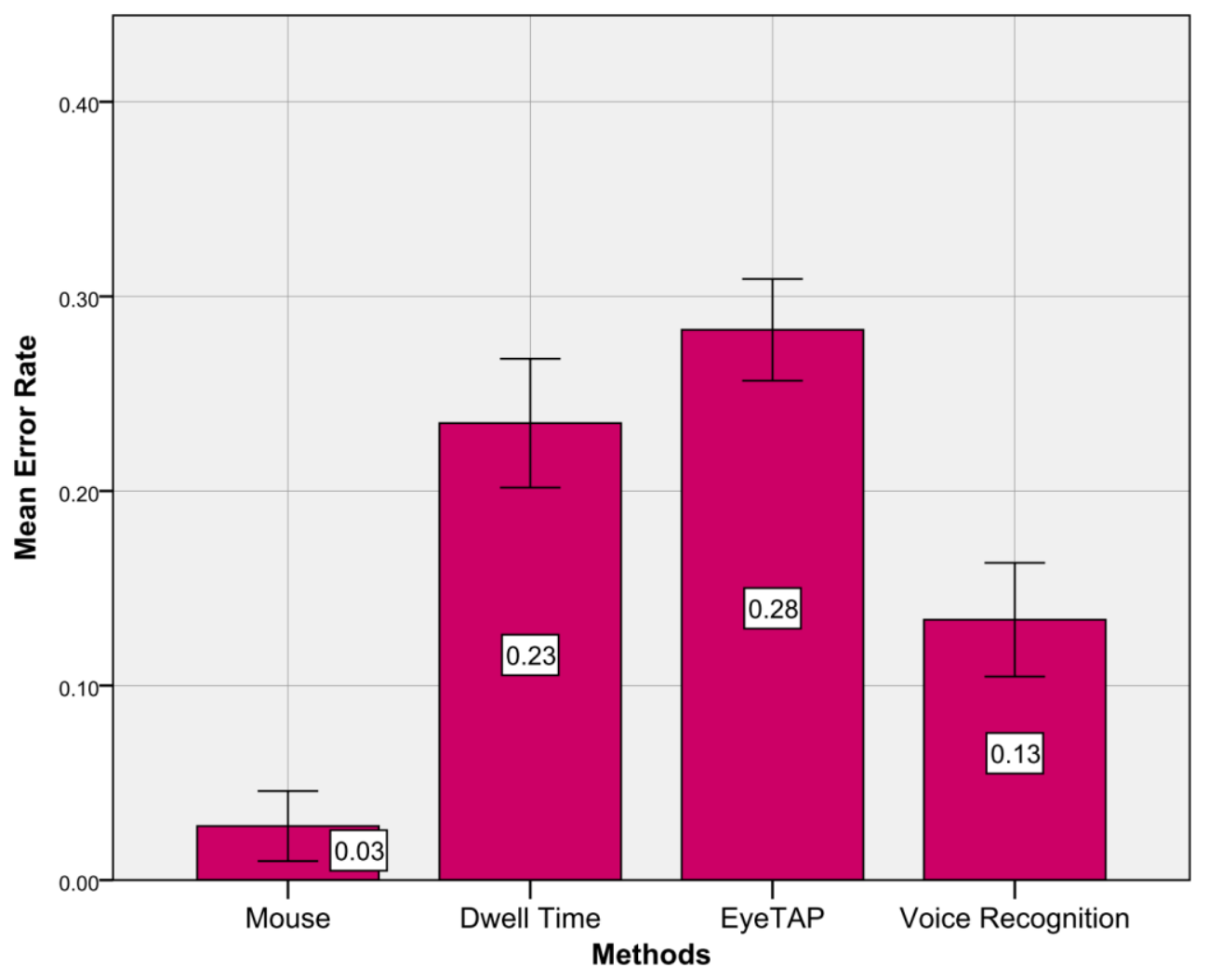}
	\caption{The calculated error rates per method for the circle-shaped test ($p < .001$).}
	\label{fig:circle_error_rates}
\end{figure}

We found a significant effect of the interaction method on error rates (F(3,96)=18.25, \textit{p} < .001). A posthoc Bonferroni comparison test showed a significant difference between mouse ($M=0.02~errors$, $SE=0.01~errors$), dwell-time ($M=0.23~errors$, $SE=0.03~errors$), voice recognition ($M=0.13~errors$, $SE=0.02~errors$) and EyeTAP ($M=0.28~errors$, $SE=0.02~errors$). Voice recognition ($M=0.13~errors$, $SE=0.02~errors$) reached the lowest error rate among eye tracking methods, however, there is no statistical difference between dwell-time ($M=0.23~errors$, $SE=0.03~errors$) and EyeTAP ($M=0.28~errors$, $SE=0.02~errors$). Figure \ref{fig:circle_error_rates} illustrates the calculated error rates for the circle-shaped test.

\subsection{EyeTAP rating by users}
\begin{figure}[htbp]
	\centering
		\includegraphics[width=0.47 \textwidth]{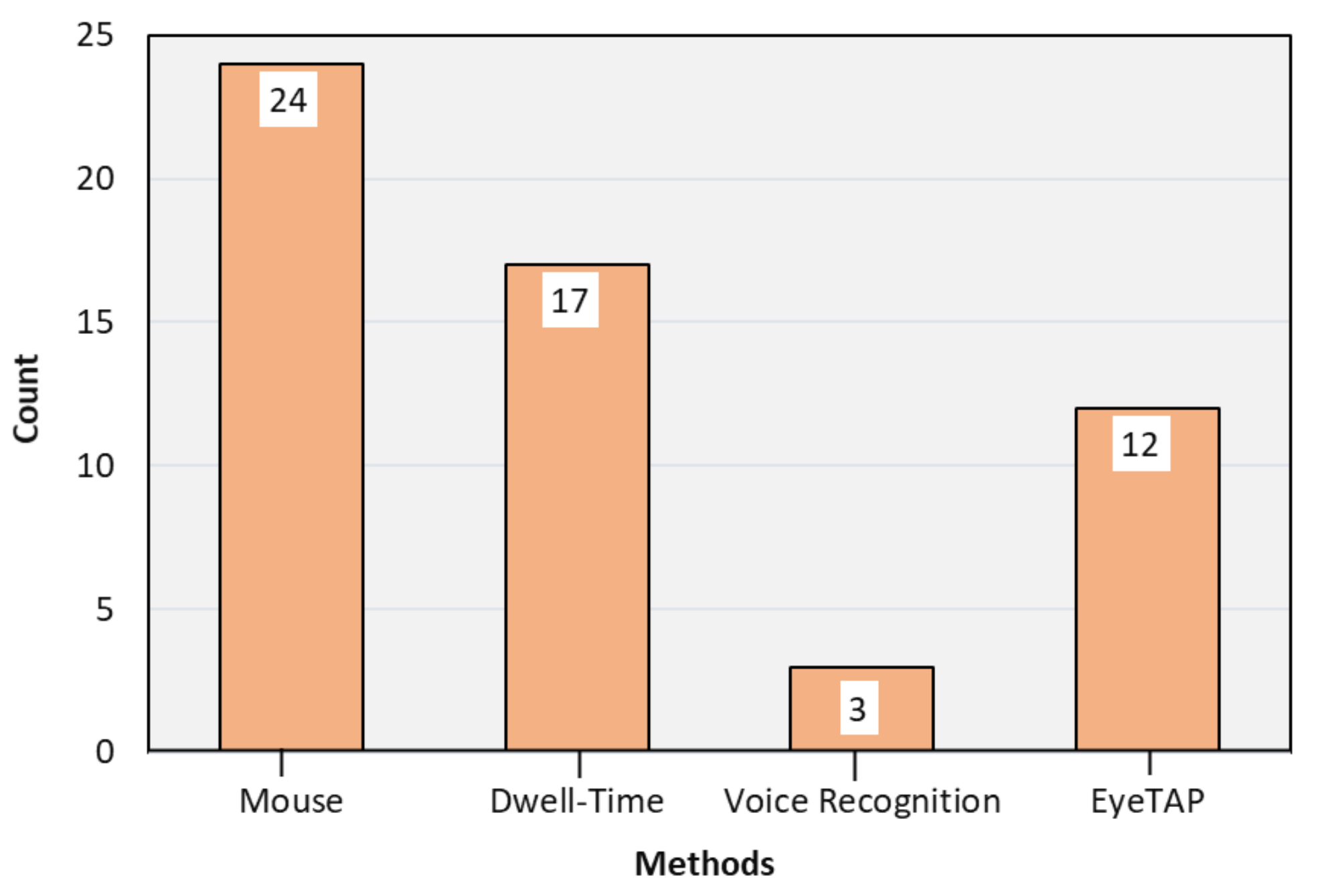}
	\caption{The recorded users' multiple choice of interaction techniques among 33 participants.}
	\label{fig:techniques_rating_chart}
\end{figure}

We asked participants to evaluate the overall performance of EyeTAP in the post-test questionnaire on a scale from 1 (worst) to 5 (best). EyeTAP reached the average rate of 3.64 ($SD=0.99$) by 33 users. In addition, Users were asked to select multiple interaction techniques. Figure \ref{fig:techniques_rating_chart} illustrates the popular interaction techniques by users obtained from the post-test questionnaire. EyeTAP reached the second desired eye tracking technique. 

\subsection{NASA TLX scores}
\begin{figure*}[htbp]
	\centering
		\includegraphics[width=.99 \textwidth]{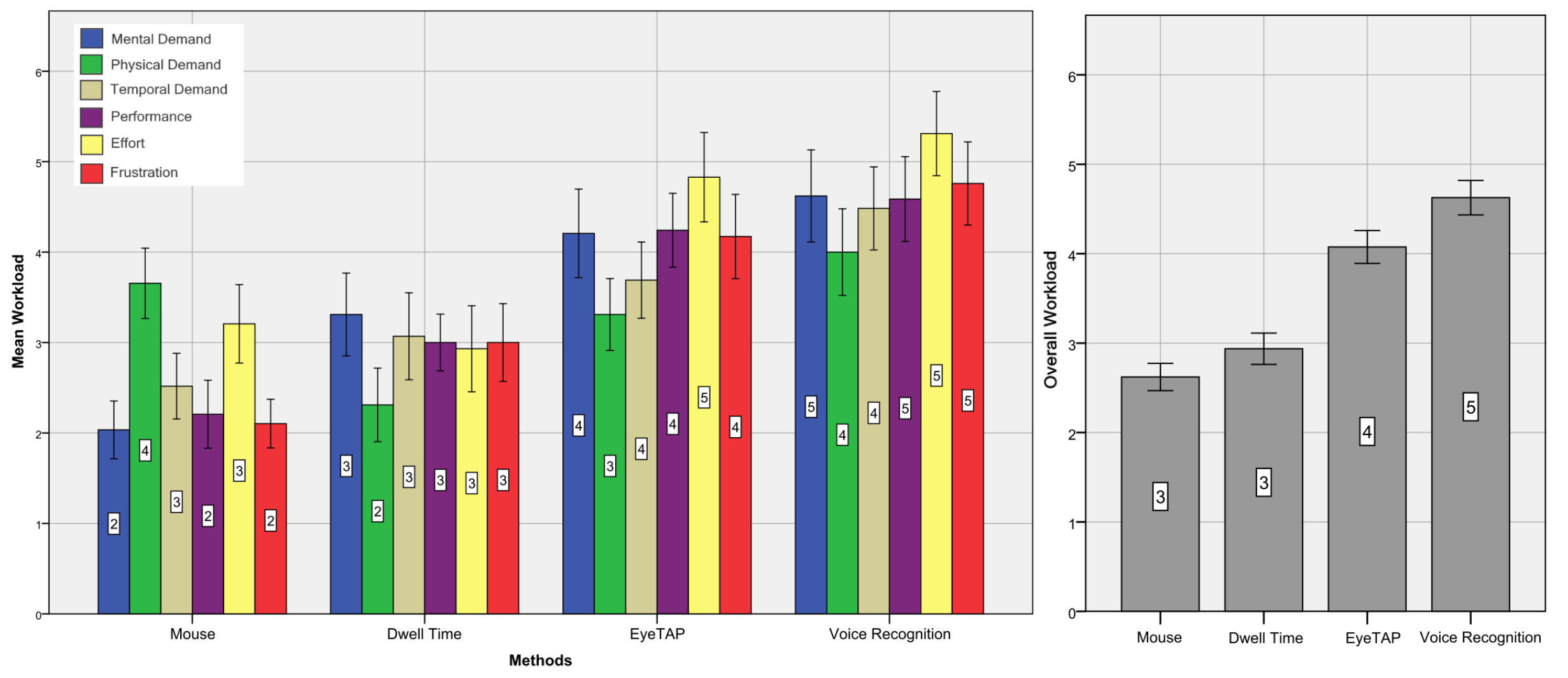}
	\caption{The NASA TLX scores for the interaction methods. (Left) Comparison of each method based on different scales. (Right) The overall mean workload of tested interaction methods.}
	\label{fig:nasa_tlx_chart}
\end{figure*}

Figure \ref{fig:nasa_tlx_chart} shows the NASA TLX scores for all interaction methods obtained during the user study. The overall workload is the average of scale values since we assume all scales equally important and therefore eliminated the weighting calculations to apply a simplified version \cite{nasa_tlx_20} of the basic NASA TLX ratings \cite{nasa_tlx}. According to our findings, the dwell-time method has the lowest workload among other eye tracking techniques. However, EyeTAP shows relatively lower workload compared to voice recognition technique.

\section{Discussion}
Regarding the experiments with the reviewed Midas touch solutions, we found several benefits and disadvantages of each method. We discuss each method individually.

\subsection{Voice Recognition}
This interaction method showed relatively acceptable results but suffers from some limitations. In general, a voice recognition engine depends on the user's voice, gender, language, and accent. Additionally, it is not applicable to users with speech impediments. Another drawback is the need of prior training samples to detect words correctly. Furthermore, similar words may lead to false recognition as we experienced during our user study. The quality of the microphone and its distance to the user is also another factor to be considered for this kind of interaction. Regarding the accuracy of recognition, the choice of recognition software plays an important role. Finally, speaking out loud may not be suitable in certain working environments. 

In general, voice recognition presented some challenges for the users in terms of wrongly recognized words, need for action word repetition, and delay between input and feedback. The subjects' rating of this technique was very low (9.1\%) in our user study. Voice recognition showed the highest completion time in the matrix-based test and highest movement time in the circle-shaped test and reached the highest cognitive workload among all interaction techniques. However, voice recognition showed the lowest error rates in both Fitts' study experiments and reached the lowest distance to target (highest selection accuracy) among other eye tracking techniques.

\subsection{Dwell-Time processing}
Dwell-time method showed the fastest completion time in the matrix-based test, and fastest movement time and highest throughput in both Fitts' experiments due to the low amount of activation time (500 ms). In addition, it reached the lowest amount of cognitive workload. However, it showed the highest error rates in the ribbon-shaped test and with EyeTAP in the circle-shaped test. Moreover, some users complained about eye fatigue after a while during test sessions. 


\subsection{EyeTAP}
We found several benefits of using EyeTAP in comparison to the other interaction techniques. First of all, it has no dependent features, rather it requires only an acoustic pulse (making sound with mouth) near a microphone to send a signal. In fact, the output of EyeTAP in a noisy environment (up to 70dB) can appear deterministic after a number of repetitions. According to the results of our study, it achieved faster completion time in the matrix-based test, and faster movement time in the circle-shaped experiment than voice recognition. In addition, it showed a similar path cost (pointer footprint on display) with the other eye tracking techniques. It also achieved lower cognitive workload in comparison to the voice recognition technique. Furthermore, EyeTAP was the popular choice of interaction (36.4\%) compared to voice recognition (9.1\%). However, EyeTAP showed relatively lower accuracy and higher error rates than voice recognition, since most users had no prior experiences with this kind of interaction. The performance of EyeTAP can be improved with more training. 

In general, EyeTAP is simple, integrates well into existing user interfaces, and allows for easy and accurate point-and-select interaction because it separates the actions of \textit{pointing} and \textit{selecting} to two different modalities while relaxing the requirement for accurate voice recognition. The results of our user study demonstrate that EyeTAP is a feasible alternative interaction technique. Moreover, it is a robust and effective solution to the Midas touch problem for eye tracking platforms and can be regarded as an alternative to voice recognition technique.


\section{Conclusion and Future Work}

In this paper, we proposed EyeTAP (Eye tracking point-and-select by Targeted Acoustic Pulse), an eye-tracking interface that addresses the Midas touch problem with acoustic input detection capabilities. EyeTAP allows for accurate and effective interaction without the need for extra equipment or user interface design for gaze-based interactions. The performance of the prototype was measured in two independent user studies with 33 participants based on eight criteria: (1) \textit{completion time}, (2) \textit{path cost of target selection}, (3) \textit{error rate}, (4) \textit{error locations on screen}, (5) \textit{accuracy of target selection}, (6) \textit{movement time}, (7) \textit{throughput}, and (8) \textit{cognitive workload}. 

The results of our user studies showed that the dwell-time method outperformed other eye tracking techniques, including EyeTAP on most criteria. At the same time we found that EyeTAP, in comparison to to the other tested methods is a competitive and a promising solution and provides a faster task completion time, faster movement time and lower workload than voice recognition. In addition, EyeTAP showed similar performance compared to the dwell-time method and lower error rate in the ribbon-shaped experiment. 

Moreover, our study showed that eye tracking has a lower footprint on the screen compared to a mouse pointer in time scale. Additionally, we confirmed that center regions towards the right and bottom side of the screen are more error prone than the left and top sides. Additionally, we developed two user tests that would be effective in studying different target selection for gaze-based interaction techniques. 

Although we only developed the left mouse click event, EyeTAP demonstrates a completely hands-free or touchless alternative to mouse interaction for users with disabilities and users who need to avoid physical contact with input devices considering their workplace or situation. Thus, we believe EyeTAP can be regarded as a competitive technique to both dwell-time and voice recognition. In future work, we will apply the EyeTAP technique on AR/VR headsets to measure its usability in different case scenarios.


\bibliographystyle{plainnat}
\bibliography{bibliography}

\begin{thebibliography}{38}
\providecommand{\natexlab}[1]{#1}
\providecommand{\url}[1]{\texttt{#1}}
\expandafter\ifx\csname urlstyle\endcsname\relax
  \providecommand{\doi}[1]{doi: #1}\else
  \providecommand{\doi}{doi: \begingroup \urlstyle{rm}\Url}\fi

\bibitem[B\^{a}ce et~al.(2016)B\^{a}ce, Lepp\"{a}nen, de~Gomez, and
  Gomez]{GazeGesture2016}
Mihai B\^{a}ce, Teemu Lepp\"{a}nen, David~Gil de~Gomez, and Argenis~Ramirez
  Gomez.
\newblock ubigaze: Ubiquitous augmented reality messaging using gaze gestures.
\newblock In \emph{SIGGRAPH ASIA 2016 Mobile Graphics and Interactive
  Applications}, SA '16, pages 11:1--11:5, New York, NY, USA, 2016. ACM.
\newblock ISBN 978-1-4503-4551-4.
\newblock \doi{10.1145/2999508.2999530}.
\newblock URL \url{http://doi.acm.org/10.1145/2999508.2999530}.

\bibitem[Barnes(2012)]{Barnes2012}
Graham~R. Barnes.
\newblock Rapid learning of pursuit target motion trajectories revealed by
  responses to randomized transient sinusoids.
\newblock \emph{Journal of Eye Movement Research}, 5\penalty0 (3), 2012.
\newblock ISSN 1995-8692.
\newblock URL \url{https://bop.unibe.ch/JEMR/article/view/2337}.

\bibitem[Bednarik et~al.(2009)Bednarik, Gowases, and
  Tukiainen]{Bednarik_Gowases_Tukiainen_2009}
Roman Bednarik, Tersia Gowases, and Markku Tukiainen.
\newblock Gaze interaction enhances problem solving: Effects of dwell-time
  based, gaze-augmented, and mouse interaction on problem-solving strategies
  and user experience.
\newblock \emph{Journal of Eye Movement Research}, 3\penalty0 (1), Aug. 2009.
\newblock \doi{10.16910/jemr.3.1.3}.
\newblock URL \url{https://bop.unibe.ch/JEMR/article/view/2287}.

\bibitem[Bellman and Kalaba(1959)]{1104847}
R.~Bellman and R.~Kalaba.
\newblock On adaptive control processes.
\newblock \emph{IRE Transactions on Automatic Control}, 4\penalty0
  (2):\penalty0 1--9, November 1959.
\newblock ISSN 0096-199X.
\newblock \doi{10.1109/TAC.1959.1104847}.

\bibitem[Biswas and Langdon(2015)]{JoystickSwitch}
Pradipta Biswas and Pat Langdon.
\newblock Multimodal intelligent eye-gaze tracking system.
\newblock \emph{International Journal of Human-Computer Interaction},
  31\penalty0 (4):\penalty0 277--294, 2015.

\bibitem[Dey et~al.(2004)Dey, Hamid, Beckmann, Li, and Hsu]{Cappella}
Anind~K. Dey, Raffay Hamid, Chris Beckmann, Ian Li, and Daniel Hsu.
\newblock A cappella: Programming by demonstration of context-aware
  applications.
\newblock In \emph{Proceedings of the SIGCHI Conference on Human Factors in
  Computing Systems}, CHI '04, pages 33--40, New York, NY, USA, 2004. ACM.
\newblock ISBN 1-58113-702-8.
\newblock \doi{10.1145/985692.985697}.
\newblock URL \url{http://doi.acm.org/10.1145/985692.985697}.

\bibitem[Drewes and Schmidt(2007)]{GazeGesture2007}
Heiko Drewes and Albrecht Schmidt.
\newblock Interacting with the computer using gaze gestures.
\newblock In C{\'e}cilia Baranauskas, Philippe Palanque, Julio Abascal, and
  Simone Diniz~Junqueira Barbosa, editors, \emph{Human-Computer Interaction --
  INTERACT 2007}, pages 475--488, Berlin, Heidelberg, 2007. Springer Berlin
  Heidelberg.
\newblock ISBN 978-3-540-74800-7.

\bibitem[Esteves et~al.(2015)Esteves, Velloso, Bulling, and
  Gellersen]{OrbitsSmartWatch}
Augusto Esteves, Eduardo Velloso, Andreas Bulling, and Hans Gellersen.
\newblock Orbits: Gaze interaction for smart watches using smooth pursuit eye
  movements.
\newblock In \emph{Proceedings of the 28th Annual ACM Symposium on User
  Interface Software \&\#38; Technology}, UIST '15, pages 457--466, New York,
  NY, USA, 2015. ACM.
\newblock ISBN 978-1-4503-3779-3.
\newblock \doi{10.1145/2807442.2807499}.
\newblock URL \url{http://doi.acm.org/10.1145/2807442.2807499}.

\bibitem[Feit et~al.(2017)Feit, Williams, Toledo, Paradiso, Kulkarni, Kane, and
  Morris]{feit2017toward}
Anna~Maria Feit, Shane Williams, Arturo Toledo, Ann Paradiso, Harish Kulkarni,
  Shaun Kane, and Meredith~Ringel Morris.
\newblock Toward everyday gaze input: Accuracy and precision of eye tracking
  and implications for design.
\newblock In \emph{Proceedings of the 2017 Chi conference on human factors in
  computing systems}, pages 1118--1130. ACM, 2017.

\bibitem[Group(1986)]{nasa_tlx}
NASA Human Performance~Research Group.
\newblock Nasa task load index (tlx) paper and pencil package, 1986.
\newblock URL
  \url{https://humansystems.arc.nasa.gov/groups/TLX/downloads/TLX.pdf}.

\bibitem[Hart(2006)]{nasa_tlx_20}
Sandra~G. Hart.
\newblock Nasa-task load index (nasa-tlx); 20 years later.
\newblock \emph{Proceedings of the Human Factors and Ergonomics Society Annual
  Meeting}, 50\penalty0 (9):\penalty0 904--908, 2006.
\newblock \doi{10.1177/154193120605000909}.
\newblock URL \url{https://doi.org/10.1177/154193120605000909}.

\bibitem[Hartmann et~al.(2007)Hartmann, Abdulla, Mittal, and
  Klemmer]{DirectManipulationPatternRecognition}
Bj\"{o}rn Hartmann, Leith Abdulla, Manas Mittal, and Scott~R. Klemmer.
\newblock Authoring sensor-based interactions by demonstration with direct
  manipulation and pattern recognition.
\newblock In \emph{Proceedings of the SIGCHI Conference on Human Factors in
  Computing Systems}, CHI '07, pages 145--154, New York, NY, USA, 2007. ACM.
\newblock ISBN 978-1-59593-593-9.
\newblock \doi{10.1145/1240624.1240646}.
\newblock URL \url{http://doi.acm.org/10.1145/1240624.1240646}.

\bibitem[Hyrskykari et~al.(2012)Hyrskykari, Istance, and
  Vickers]{GazeGesture2012}
Aulikki Hyrskykari, Howell Istance, and Stephen Vickers.
\newblock Gaze gestures or dwell-based interaction?
\newblock In \emph{Proceedings of the Symposium on Eye Tracking Research and
  Applications}, ETRA '12, pages 229--232, New York, NY, USA, 2012. ACM.
\newblock ISBN 978-1-4503-1221-9.
\newblock \doi{10.1145/2168556.2168602}.
\newblock URL \url{http://doi.acm.org/10.1145/2168556.2168602}.

\bibitem[Istance et~al.(2010)Istance, Hyrskykari, Immonen, Mansikkamaa, and
  Vickers]{GazeGesture2010}
Howell Istance, Aulikki Hyrskykari, Lauri Immonen, Santtu Mansikkamaa, and
  Stephen Vickers.
\newblock Designing gaze gestures for gaming: An investigation of performance.
\newblock In \emph{Proceedings of the 2010 Symposium on Eye-Tracking Research
  \&\#38; Applications}, ETRA '10, pages 323--330, New York, NY, USA, 2010.
  ACM.
\newblock ISBN 978-1-60558-994-7.
\newblock \doi{10.1145/1743666.1743740}.
\newblock URL \url{http://doi.acm.org/10.1145/1743666.1743740}.

\bibitem[Jacob(1990)]{MidasTouchDefinition}
Robert J.~K. Jacob.
\newblock What you look at is what you get: Eye movement-based interaction
  techniques.
\newblock In \emph{Proceedings of the SIGCHI Conference on Human Factors in
  Computing Systems}, CHI '90, pages 11--18, New York, NY, USA, 1990. ACM.
\newblock ISBN 0-201-50932-6.
\newblock \doi{10.1145/97243.97246}.
\newblock URL \url{http://doi.acm.org/10.1145/97243.97246}.

\bibitem[MacKenzie(2012)]{mackenzie2012evaluating}
I~Scott MacKenzie.
\newblock Evaluating eye tracking systems for computer input.
\newblock In \emph{Gaze interaction and applications of eye tracking: Advances
  in assistive technologies}, pages 205--225. IGI Global, 2012.

\bibitem[Majaranta et~al.(2006)Majaranta, MacKenzie, Aula, and
  R{\"a}ih{\"a}]{majaranta2006effects}
P{\"a}ivi Majaranta, I~Scott MacKenzie, Anne Aula, and Kari-Jouko
  R{\"a}ih{\"a}.
\newblock Effects of feedback and dwell time on eye typing speed and accuracy.
\newblock \emph{Universal Access in the Information Society}, 5\penalty0
  (2):\penalty0 199--208, 2006.

\bibitem[Meena et~al.(2017)Meena, Cecotti, Wong-Lin, and
  Prasad]{WheelchairSwitch}
Y.~K. Meena, H.~Cecotti, K.~Wong-Lin, and G.~Prasad.
\newblock A multimodal interface to resolve the midas-touch problem in gaze
  controlled wheelchair.
\newblock In \emph{2017 39th Annual International Conference of the IEEE
  Engineering in Medicine and Biology Society (EMBC)}, pages 905--908. IEEE,
  July 2017.
\newblock \doi{10.1109/EMBC.2017.8036971}.

\bibitem[Miniotas et~al.(2004)Miniotas, \v{S}pakov, and
  MacKenzie]{JitterDefinition}
Darius Miniotas, Oleg \v{S}pakov, and I.~Scott MacKenzie.
\newblock Eye gaze interaction with expanding targets.
\newblock In \emph{CHI '04 Extended Abstracts on Human Factors in Computing
  Systems}, CHI EA '04, pages 1255--1258, New York, NY, USA, 2004. ACM.
\newblock ISBN 1-58113-703-6.
\newblock \doi{10.1145/985921.986037}.
\newblock URL \url{http://doi.acm.org/10.1145/985921.986037}.

\bibitem[Myers et~al.(1980)Myers, Rabiner, and Rosenberg]{1163491}
C.~Myers, L.~Rabiner, and A.~Rosenberg.
\newblock Performance tradeoffs in dynamic time warping algorithms for isolated
  word recognition.
\newblock \emph{IEEE Transactions on Acoustics, Speech, and Signal Processing},
  28\penalty0 (6):\penalty0 623--635, December 1980.
\newblock ISSN 0096-3518.
\newblock \doi{10.1109/TASSP.1980.1163491}.

\bibitem[of~Encyclopaedia~Britannica(2018)]{morse_code}
The~Editors of~Encyclopaedia~Britannica.
\newblock Morse code, March 2018.
\newblock URL \url{https://www.britannica.com/topic/Morse-Code}.
\newblock [Online; accessed September 14, 2018].

\bibitem[Patel and Abowd(2007)]{Blui}
Shwetak~N. Patel and Gregory~D. Abowd.
\newblock Blui: Low-cost localized blowable user interfaces.
\newblock In \emph{Proceedings of the 20th Annual ACM Symposium on User
  Interface Software and Technology}, UIST '07, pages 217--220, New York, NY,
  USA, 2007. ACM.
\newblock ISBN 978-1-59593-679-0.
\newblock \doi{10.1145/1294211.1294250}.
\newblock URL \url{http://doi.acm.org/10.1145/1294211.1294250}.

\bibitem[Pfeuffer and Gellersen(2016)]{GazeAndTouch01}
Ken Pfeuffer and Hans Gellersen.
\newblock Gaze and touch interaction on tablets.
\newblock In \emph{Proceedings of the 29th Annual Symposium on User Interface
  Software and Technology}, UIST '16, pages 301--311, New York, NY, USA, 2016.
  ACM.
\newblock ISBN 978-1-4503-4189-9.
\newblock \doi{10.1145/2984511.2984514}.
\newblock URL \url{http://doi.acm.org/10.1145/2984511.2984514}.

\bibitem[Pfeuffer et~al.(2016)Pfeuffer, Alexander, and
  Gellersen]{GazeAndTouch02}
Ken Pfeuffer, Jason Alexander, and Hans Gellersen.
\newblock Partially-indirect bimanual input with gaze, pen, and touch for pan,
  zoom, and ink interaction.
\newblock In \emph{Proceedings of the 2016 CHI Conference on Human Factors in
  Computing Systems}, CHI '16, pages 2845--2856, New York, NY, USA, 2016. ACM.
\newblock ISBN 978-1-4503-3362-7.
\newblock \doi{10.1145/2858036.2858201}.
\newblock URL \url{http://doi.acm.org/10.1145/2858036.2858201}.

\bibitem[Pi and Shi(2017)]{ProbabilisticDwellTime}
J.~Pi and B.~E. Shi.
\newblock Probabilistic adjustment of dwell time for eye typing.
\newblock In \emph{2017 10th International Conference on Human System
  Interactions (HSI)}, pages 251--257. IEEE, July 2017.
\newblock \doi{10.1109/HSI.2017.8005041}.

\bibitem[Rajanna and Hammond(2018)]{FootSwitch}
Vijay Rajanna and Tracy Hammond.
\newblock A gaze-assisted multimodal approach to rich and accessible
  human-computer interaction.
\newblock \emph{CoRR}, abs/1803.04713, 2018.
\newblock URL \url{http://arxiv.org/abs/1803.04713}.

\bibitem[Rozado et~al.(2017)Rozado, Niu, and Lochner]{FaceGestures}
David Rozado, Jason Niu, and Martin Lochner.
\newblock Fast human-computer interaction by combining gaze pointing and face
  gestures.
\newblock \emph{ACM Transactions on Accessible Computing (TACCESS)},
  10\penalty0 (3):\penalty0 10, 2017.

\bibitem[Sakoe and Chiba(1978)]{1163055}
H.~Sakoe and S.~Chiba.
\newblock Dynamic programming algorithm optimization for spoken word
  recognition.
\newblock \emph{IEEE Transactions on Acoustics, Speech, and Signal Processing},
  26\penalty0 (1):\penalty0 43--49, February 1978.
\newblock ISSN 0096-3518.
\newblock \doi{10.1109/TASSP.1978.1163055}.

\bibitem[Schenk et~al.(2017)Schenk, Dreiser, Rigoll, and Dorr]{GazeEverywhere}
Simon Schenk, Marc Dreiser, Gerhard Rigoll, and Michael Dorr.
\newblock Gazeeverywhere: Enabling gaze-only user interaction on an unmodified
  desktop pc in everyday scenarios.
\newblock In \emph{Proceedings of the 2017 CHI Conference on Human Factors in
  Computing Systems}, CHI '17, pages 3034--3044, New York, NY, USA, 2017. ACM.
\newblock ISBN 978-1-4503-4655-9.
\newblock \doi{10.1145/3025453.3025455}.
\newblock URL \url{http://doi.acm.org/10.1145/3025453.3025455}.

\bibitem[Sibert and Jacob(2000)]{SibertJacob2000}
Linda~E. Sibert and Robert J.~K. Jacob.
\newblock Evaluation of eye gaze interaction.
\newblock In \emph{Proceedings of the SIGCHI Conference on Human Factors in
  Computing Systems}, CHI ’00, page 281–288, New York, NY, USA, 2000.
  Association for Computing Machinery.
\newblock ISBN 1581132166.
\newblock \doi{10.1145/332040.332445}.
\newblock URL \url{https://doi.org/10.1145/332040.332445}.

\bibitem[Sidorakis et~al.(2015)Sidorakis, Koulieris, and
  Mania]{BinocularEyeTracking}
N.~Sidorakis, G.~A. Koulieris, and K.~Mania.
\newblock Binocular eye-tracking for the control of a 3d immersive multimedia
  user interface.
\newblock In \emph{2015 IEEE 1st Workshop on Everyday Virtual Reality (WEVR)},
  pages 15--18. IEEE, March 2015.
\newblock \doi{10.1109/WEVR.2015.7151689}.

\bibitem[{\v{S}}pakov and Miniotas(2004)]{vspakov2004line}
Oleg {\v{S}}pakov and Darius Miniotas.
\newblock On-line adjustment of dwell time for target selection by gaze.
\newblock In \emph{Proceedings of the third Nordic conference on Human-computer
  interaction}, pages 203--206. ACM, 2004.

\bibitem[Velichkovsky et~al.(2014)Velichkovsky, Rumyantsev, and
  Morozov]{FocalFixations}
Boris~B Velichkovsky, Mikhail~A Rumyantsev, and Mikhail~A Morozov.
\newblock New solution to the midas touch problem: Identification of visual
  commands via extraction of focal fixations.
\newblock \emph{Procedia Computer Science}, 39:\penalty0 75--82, 2014.

\bibitem[Velloso et~al.(2016)Velloso, Wirth, Weichel, Esteves, and
  Gellersen]{AmbiGaze}
Eduardo Velloso, Markus Wirth, Christian Weichel, Augusto Esteves, and Hans
  Gellersen.
\newblock Ambigaze: Direct control of ambient devices by gaze.
\newblock In \emph{Proceedings of the 2016 ACM Conference on Designing
  Interactive Systems}, DIS '16, pages 812--817, New York, NY, USA, 2016. ACM.
\newblock ISBN 978-1-4503-4031-1.
\newblock \doi{10.1145/2901790.2901867}.
\newblock URL \url{http://doi.acm.org/10.1145/2901790.2901867}.

\bibitem[Vertegaal(2008)]{10.1145/1452392.1452443}
Roel Vertegaal.
\newblock A fitts law comparison of eye tracking and manual input in the
  selection of visual targets.
\newblock In \emph{Proceedings of the 10th International Conference on
  Multimodal Interfaces}, ICMI ’08, page 241–248, New York, NY, USA, 2008.
  Association for Computing Machinery.
\newblock ISBN 9781605581989.
\newblock \doi{10.1145/1452392.1452443}.
\newblock URL \url{https://doi.org/10.1145/1452392.1452443}.

\bibitem[Vidal et~al.(2013)Vidal, Bulling, and Gellersen]{Pursuits}
M{\'e}lodie Vidal, Andreas Bulling, and Hans Gellersen.
\newblock Pursuits: Spontaneous interaction with displays based on smooth
  pursuit eye movement and moving targets.
\newblock In \emph{Proceedings of the 2013 ACM International Joint Conference
  on Pervasive and Ubiquitous Computing}, UbiComp '13, pages 439--448, New
  York, NY, USA, 2013. ACM.
\newblock ISBN 978-1-4503-1770-2.
\newblock \doi{10.1145/2493432.2493477}.
\newblock URL \url{http://doi.acm.org/10.1145/2493432.2493477}.

\bibitem[\v{S}pakov and Miniotas(2004)]{10.1145/1028014.1028045}
Oleg \v{S}pakov and Darius Miniotas.
\newblock On-line adjustment of dwell time for target selection by gaze.
\newblock In \emph{Proceedings of the Third Nordic Conference on Human-Computer
  Interaction}, NordiCHI ’04, page 203–206, New York, NY, USA, 2004.
  Association for Computing Machinery.
\newblock ISBN 1581138571.
\newblock \doi{10.1145/1028014.1028045}.
\newblock URL \url{https://doi.org/10.1145/1028014.1028045}.

\bibitem[Wobbrock et~al.(2011)Wobbrock, Shinohara, and
  Jansen]{wobbrock2011effects}
Jacob~O Wobbrock, Kristen Shinohara, and Alex Jansen.
\newblock The effects of task dimensionality, endpoint deviation, throughput
  calculation, and experiment design on pointing measures and models.
\newblock In \emph{Proceedings of the SIGCHI Conference on Human Factors in
  Computing Systems}, pages 1639--1648. ACM, 2011.

\end{thebibliography}

\end{document}